\documentclass[twocolumn,aps,prc,amsmath,amssymb,graphicx,longbibliography]{revtex4}

\usepackage{epsfig}
\usepackage{epsf}

\usepackage{bm}

\newcommand{\bx}{{\bf x}}
\newcommand{\bk}{{\bf k}}

\newcommand{\vm}{{\bf m}}
\newcommand{\vn}{{\bf n}}
\newcommand{\vM}{{\bf M}}
\newcommand{\Heff}{{\bf H}_{\rm eff}}
\newcommand{\Hext}{{\bf H}_{\rm ext}}

\begin{document}

\title{Hannay Angles in Magnetic Dynamics}

\author{A. R\"uckriegel}
\affiliation{Institute for Theoretical Physics, Utrecht
	University, Princetonplein 5, 3584 CC Utrecht, The Netherlands}

\author{R. A. Duine}

\affiliation{Institute for Theoretical Physics, Utrecht
	University, Princetonplein 5, 3584 CC Utrecht, The Netherlands}

\affiliation{Department of Applied Physics, Eindhoven University of Technology, P.O. Box 513, 5600 MB Eindhoven, The Netherlands}

\date{\today{}}

\begin{abstract}
We consider, within the framework developed by Hannay for classical integrable systems [Journal of Physics A: Mathematical and General {\bf 18}, 221 (1985)], the geometric phases that occur in semi-classical magnetic dynamics. Such geometric phases are generically referred to as Hannay angles, and, in the context of magnetic dynamics, may arise as a result of both adiabatically-varying ellipticity and axis of magnetization precession. We elucidate both effects and their interplay for single-domain magnetic dynamics within a simple model with time-dependent anisotropies and external field. Subsequently, we consider spin waves and rederive, from our classical approach, some known results on what is commonly referred to as the magnon Berry phase. As an aside, these results are used to give an interpretation for geometric phases that occur in superfluids. Finally, we develop a Green's function formalism for elliptical magnons. Within this formalism, we consider magnon transport in a mesoscopic ring and show how it is influenced by interference effects that are tuned by the Hannay angle that results from a varying ellipticity. Our results may inform the field of magnonics that seeks to utilize spin waves in applications.
\end{abstract}


\maketitle

\tableofcontents{}

\section{Introduction}
\label{sec:intro}

Consider the following exercise, that could have featured as part of a classical-mechanics course that you took: a particle is confined to move freely on an ellipse and set into motion with some given energy. As the particle moves on the ellipse, the ellipse itself is rotated once while keeping the plane in which the particle moves the same. The angular velocity of the ellipse's rotation is very small compared to the angular velocity of the particle, so that the particle completes many circuits while the ellipse rotates. How many extra circuits does the particle make if the ellipse is rotated once, as compared to the case where the ellipse is not rotated? 

The answer to this question, it turns out, does not depend on whether the ellipse is rotated with constant angular velocity, nor does it depend on how fast the ellipse is rotated --- provided it is rotated slowly enough. It solely depends on the geometry of the ellipse, or, more specifically, on the ratio of the length of its principal axes. When expressed in terms of an angle that parametrizes the position of the particle on the ellipse, the excess amount of circuits is an example of a Hannay angle \cite{Hannay_1985}. 

Such angles occur in classical confined and integrable Hamiltonian systems whenever the Hamiltonian is taken around a closed loop in parameter space. Integrability ensures the existence of adiabatic invariants, called action variables, that are conjugate to so-called angle variables. Loosely speaking, the integrability ensures periodic motion of some variable. As a result, it is parametrized by an angle and that angle may acquire a geometric contribution when the Hamiltonian is taken around a closed loop in the space of its parameters. In the example of the exercise above, the action variable conjugate to the angle that parametrizes the position of the particle on the ellipse is simply proportional to the energy of the particle. 

The Hannay angle is an example of a geometric phase. Over the past few decades, geometric phases have become part of the established vocabulary of physics \cite{chruscinski2004geometric}. An important contribution to this development was the discovery of what is now known as the Berry phase. This best-known example of a geometric phase is the phase that a quantum mechanical systems picks up when its Hamiltonian is taken around a closed loop in parameter space \cite{berry1984quantal}. The Berry phase has, for example, been important in the development of the theory of polarization \cite{Resta2007}, anomalous transport \cite{RevModPhys.82.1959, RevModPhys.82.1539}, and topological insulators \cite{RevModPhys.82.3045,RevModPhys.83.1057}. Less well-known examples of geometric phases, next to the aforementioned Hannay angle, may occur in dissipative and stochastic classical systems \cite{Sinitsyn_2009}.

In this article, we focus on the geometric phases that arise in the semi-classical spin dynamics of ordered magnetic systems. An example of such a phase is the one picked up by a spin wave as it travels through a magnetic texture with non-trivial topology. This phase is commonly referred to as a magnon Berry phase, a magnon being a quantized spin wave, and was introduced by Dugaev {\it et al.} \cite{PhysRevB.72.024456}. Its momentum-space version has been used to develop the theory of anomalous magnon transport \cite{PhysRevLett.106.197202}.

Here, we consider these geometric phases using the approach of Hannay in terms of action and angle variables. This approach does not rely on a formulation in terms of gauge fields, and provides an alternative approach. That the magnon Berry phase is actually a classical Hannay angle can be understood from the fact that it is, in  principle, directly observable by measuring the magnetization direction. This is contrary to a true quantum-mechanical Berry phase that can be observed only through interference. That the terminology ``Berry phase" is used for what are actually Hannay angles in spin systems is understandable, as they can be similar. This is illustrated by the following example: consider a quantum spin $S$ in its ground state in a Zeeman field. Taking the direction of this field around a loop on the unit sphere enclosing a solid angle $\Omega$, gives rise to the perhaps best-known example of a Berry phase, $e^{-iS\Omega}$ \cite{berry1984quantal}. In the semi-classical limit ($S \to \infty$), however, the spin undergoes circular precession around the magnetic field. Taking the field direction around the same loop as before, while the spin precesses around it, leads to an extra angle of precession, a Hannay angle, that is equal to $\Omega$. Hence, both the Berry phase and Hannay angle are, for this example, characterized by the solid angle $\Omega$. For more details on the relation between Berry phases and Hannay angles, and for a mathematically more rigorous treatment of adiabaticity in classical mechanics, the reader may consult \cite{chruscinski2004geometric}. Here, we shall not be overly concerned with mathematical rigour but will focus on physical examples instead. 

The plan of this article is as follows. In Sec.~\ref{sec:singledomain} we will introduce a toy model that allows us to discuss the Hannay angles that occur in the precession of a single-domain magnet in detail and in what is hopefully a pedagogical manner. In particular, this model allows for a detailed anatomy of the Hannay angles that occur. We shall see that there is both a Hannay angle due to the ellipticity of the precession, reminiscent of the exercise that was posed to the reader at the beginning of this introduction, and a Hannay angle due to the changing direction of field that was already briefly mentioned above in terms of the solid angle $\Omega$. Using the understanding of single-domain precessional dynamics, we consider the geometric phase picked up by single coherent spin wave in Sec.~\ref{sec:spinwaves} and rederive some of the results of Dugaev {\it et al.} \cite{PhysRevB.72.024456} using the formulation in terms of action and angle variables. Building upon these results, we present a brief intermezzo that gives a simple interpretation of the geometric phases that occur in superfluids \cite{PhysRevLett.97.040401}. Going back to magnetic systems, we consider in Sec.~\ref{sec:magnons} a simple set-up to study how the Hannay angles due to ellipticity influence transport of incoherent (thermal) magnons. We end with a brief conclusion, discussion, and outlook.

\section{Single-domain magnetization dynamics}
\label{sec:singledomain}

We consider a single-domain ferromagnet well below the Curie temperature. Its direction of magnetization $\vm \equiv {\vM}/M_s$, with $M_s$ the saturation magnetization, obeys the Landau-Lifshitz equation \cite{LANDAU199251}
\begin{equation}
\label{eq:LLfirst}
  \frac{\partial \vm (t)}{\partial t} = - \gamma \mu_0 \vm (t) \times  \Heff (\vm (t))~,
\end{equation}
where $\gamma>0$ is minus the gyromagnetic ratio, $\mu_0$ is the vacuum permeability,  and $\Heff$ is the effective field, which is, in general, a function of $\vm$ and its spatial derivatives. The effective field is proportional to the functional derivative of the so-called micromagnetic energy functional $E[\vm]$:
\begin{equation}
\label{eq:efffieldsingledomain}
  \Heff (\vm) = - \frac{1}{\mu_0 M_s} \frac{\delta E[\vm]}{\delta \vm}~.
\end{equation}
Specific examples of the micromagnetic energy and effective fields are discussed below. Usually, one adds a Gilbert damping term $ \alpha \vm (t) \times   \partial \vm (t)/\partial t$, proportional to the dimensionless constant $\alpha \ll 1$, to the right-hand side of Eq.~(\ref{eq:LLfirst}) \cite{1353448}. The Gilbert damping term phenomenologically accounts for relaxation of the magnetic energy so that the magnetization direction eventually reaches its lowest-energy state with $\vm$ pointing along $\Heff$. Gilbert damping leads to finite time and length scales above which the geometric angles that are the focus of this article are unobservable. Apart from mentioning these time and length scales, we will, throughout this article, mostly ignore Gilbert damping and take $\alpha=0$. 

\begin{figure} 
	\includegraphics[width=9cm]{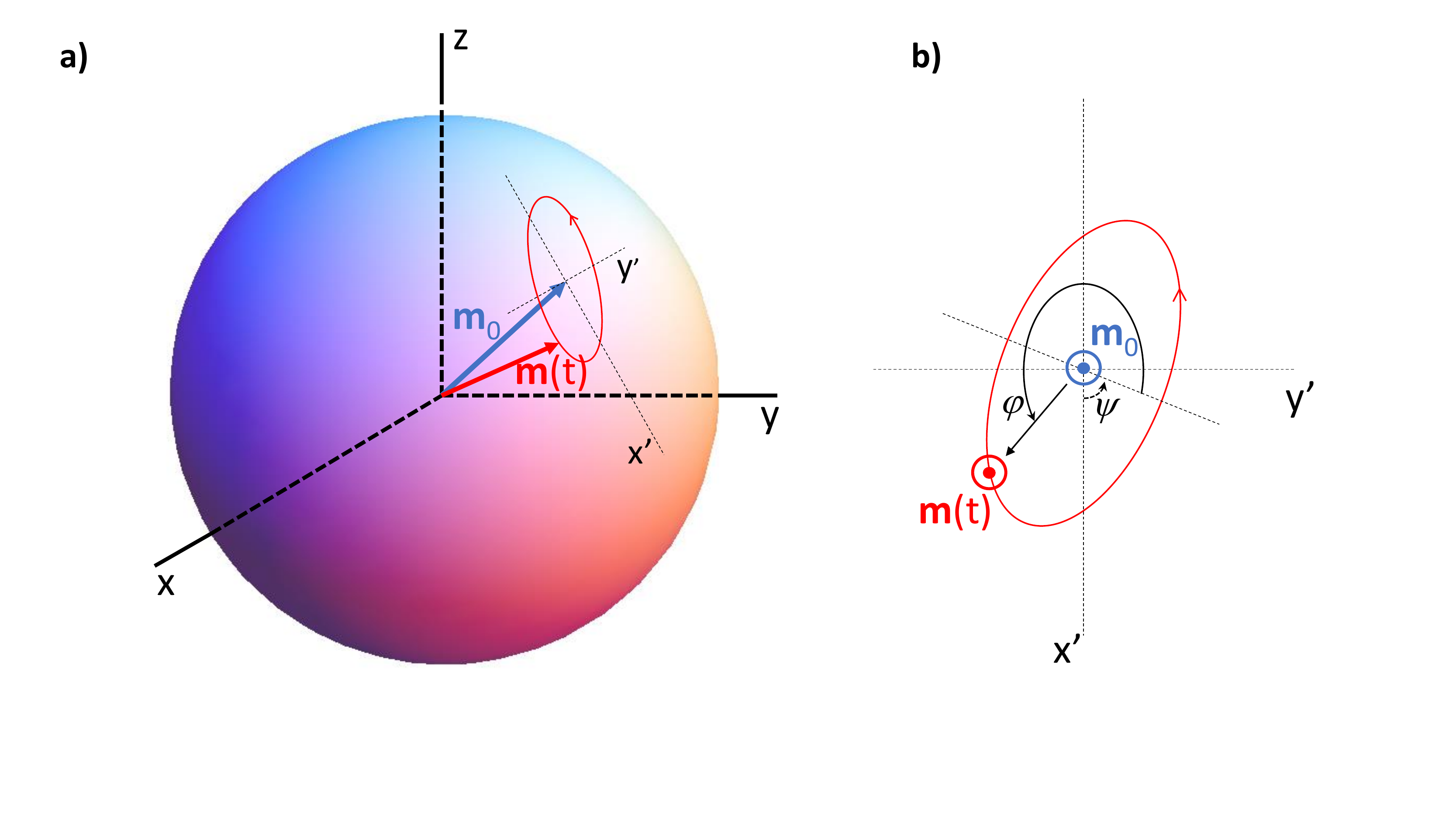} 
	\caption{a) Elliptical magnetization precession of the magnetization direction $\vm (t)$ around the equilibrium direction $\vm_0$. The coordinates $(x',y')$ label positions in the plane perpendicular to $\vm_0$. b) The angle variable $\varphi (t)$ is the angle between the projection of $\vm (t)$ onto the plane perperdicular to $\vm_0$ and one of the principal axes of the ellipse. The angle $\psi$ is the angle between one of the principal axes of the ellipse and the $x'$-direction.}
 \label{fig:illdynamics}
\end{figure}

The Landau-Lifshitz equation decribes counterclockwise precession of the magnetization direction around the effective field as illustrated in Fig.~\ref{fig:illdynamics}. The length of $\vm$ is preserved so that $\vm$ is restricted to the unit sphere. Let us first consider the case that the external magnetic field and other parameters entering $E[\vm]$, such as anisotropy constants, are time-independent. The small-angle linearized dynamics, referring to the angle between $\Heff$ and $\vm$, corresponds to an ellipse in the plane perpendicular to $\vm_0$. Here, the equibrium magnetization direction $\vm_0$ is determined by solving $\vm_0 \times \delta E[\vm_0]/\delta \vm_0 =0$ for $\vm_0$, with the restriction that $|\vm_0|=1$. Both the shape of the ellipse and the orientation of the plane in which it lies are time-independent if the micromagnetic energy is time-independent. Let us consider linearized dynamics from now on and let $\varphi$ be the angle between the magnetization projected on the plane perpendicular to $\vm_0$, with some fixed axis in the same plane (see Fig.~\ref{fig:illdynamics}). This angle constitutes an angle variable in the spirit of classical integrable systems: it is conjugate to a conserved quantity, the ``action variable" $I$. In the present case of linearized dynamics, the action variable is simply the area of the ellipse, which is proportional to the energy that the small deviation of the magnetization from its equilibrium direction $\vm_0$ costs. In general, the equations of motion for the action and angle variables are
\begin{subequations}
	\label{eq:actionangleeomgeneral}
\begin{eqnarray}
 \frac{\partial I (t)}{\partial t} &=& - \frac{\partial E}{\partial \varphi} =0~,  \\
  \frac{\partial \varphi (t)}{\partial t} &=&  \frac{\partial E}{\partial I}~,
\end{eqnarray}
\end{subequations}
where $\partial E/\partial \varphi=0$ follows from the definition of $I$ and $\varphi$.

Let us now consider a time-dependent change, starting at $t=0$, of the parameters in the energy which therefore causes the instantaneous equilibrium magnetization direction $\vm_0$ to depend on time. We now consider this change to be adiabatically slow and to result in a closed loop in the configuration space of $\vm_0$ such that $\vm_0 (0) = \vm_0 (T)$, with $T$ the time during which the adiabatic excursion takes place. We take $\vm_0$ independent of time for $t<0$ and $t>T$. For adiabatic changes in $\vm_0 (t)$, the linearized magnetization dynamics corresponds to elliptical precession around the instantaneous equilibrium magnetization direction $\vm_0 (t)$. Due to the adiabatic time-dependence of $\vm_0 (t)$ the ellipse will adiabatically change its shape whereas the plane in which it lies will change its orientation. We will consider the situation that the direction of both field and anistropy can be time-dependent, but not their magnitude, so that the energy is constant. In that case the area of the ellipse remains the same. The angle variable is not constant, however, and Hannay pointed out that there is, in addition to the dynamic contribution $\int_0^T dt \partial E/\partial I$, generically a geometric contribution to the angle variable when the system parameters are taken adiabatically around a loop in parameter space. 

In the case of magnetization dynamics this geometric contribution results from two effects. First, the orientation of the plane changes as $\vm_0$ changes. Second, the principal axes of the ellipse may change as $\vm_0$ varies, and, in particular, the ellipse may rotate. In the next subsection, we discuss a toy model that illustrates both effects. We first discuss them separately, finishing with a discussion of their interplay. 

\subsection{Toy model}
\label{subsec:atoymodel} 
\begin{table}[ptb]
	\begin{ruledtabular}
		\begin{tabular}{p{0.8\linewidth}cc}
			& Symbol\\
			\hline
			Magnetization direction & $\vm$  \\
		Equilibrium magnetization direction & $\vm_0$  \\
			Angle variable that parametrizes the precession & $\varphi$  \\
			Angle that parametrizes direction of anisotropy & $\psi$   \\
		Direction of external magnetic field & $\vn$  \\
		Angles that parametrize direction of external field as $\vn = (\sin \theta \cos \phi, \sin \theta \sin \phi, \cos \theta)$ & $\theta,\phi$  \\

		\end{tabular}
	\end{ruledtabular}
	\caption{Notation and meaning of various angles and vectors used in this article.}%
	\label{tab:angles}%
\end{table}
The micromagnetic energy for our toy model of a single-domain magnet consists of contributions due an external field $\Hext = H_{\rm ext} \vn$ in the direction ${\vn}$ and anistropies. Here, $\vn = (\sin \theta \cos \phi, \sin \theta \sin \phi, \cos \theta)$ is a unit vector that we parameterize with the angles $\theta$ and $\phi$. The angle $\theta$ corresponds to the angle between $\vn$ and the $z$-direction, whereas $\phi$ is the angle between the $x$-direction and the projection of $\vn$ onto the $x-y$-plane. (See Table~\ref{tab:angles} for an overview of the various vectors and angles.) The anisotropy that we consider corresponds to the situation that the energy cost for deviations of the magnetization directions away from $\vn$ depends on the direction of deviation. In total, the energy is
\begin{eqnarray}
\label{eq:fullenergytoymodel} 
  E[\vm] &=& - \mu_0 M_s H_{\rm ext} \vm \cdot \vn \nonumber \\
  && + \frac{K_1}{2} \left[ \vm \cdot \left( {\bf e}_x' \cos \psi  -{\bf e}_y' \sin \psi \right) \right]^2 \nonumber \\
 && + \frac{K_2}{2} \left[ \vm \cdot \left( {\bf e}_x' \sin \psi  +{\bf e}_y' \cos \psi \right) \right]^2~.
\end{eqnarray}
In this expression, $K_1>0$ and $K_2>0$ are the two constants that determine the anisotropy, with $K_1 \neq K_2$ so that the precession is, as we shall see, elliptical. The unit vectors ${\bf e}'_x$ and ${\bf e}_y'$ are perpendicular to each other and the $\vn$-direction so that they span the plane perpendicular to $\vn$. We choose ${\bf e}'_x = (\cos \theta \cos \phi, \cos \theta \sin \phi, -\sin \theta)$ and ${\bf e}'_y = (-\sin \phi, \cos \phi,0)$, so that ${\bf e}'_x$ and ${\bf e}'_y$ correspond, respectively, to the $x$ and $y$-direction when $\theta=\phi=0$. The angle $\psi$ is the angle that one of the principal axes of the ellipse of precession makes with the ${\bf e}_x'$-axis [see Fig.~\ref{fig:illdynamics}~b)].  

Consider first the situation that $\theta=\phi=\psi=0$. In this case, $\vn$ and the equilibrium magnetization direction $\vm_0$ point along the $z$-direction. We write $\vm = (m_x, m_y, 1-m_x^2/2-m_y^2/2)$, with $m_x, m_y \ll 1$, which after insertion into Eq.~(\ref{eq:fullenergytoymodel}) and expansion up to quadratic order in $m_x$ and  $m_y$ yields
\begin{equation}
\label{eq:energylinearized}
  E_0 [\vm] = \frac{\mu_0 M_s H_{\rm ext}}{2} \left( m_x^2 +  m_y^2 \right) + \frac{K_1}{2}  m_x^2+\frac{K_2}{2}  m_y^2~,
\end{equation}
where we have omitted a constant that is irrelevant for our purposes. The contours of constant energy corresponds to ellipses in the $(m_x, m_y)$-plane of which the principal axes are aligned with the $m_x$ and $m_y$-direction. The ratio between the length of the principle axes in these respective directions is $\sqrt{(\mu_0 M_s H_{\rm ext}+K_1)/(\mu_0 M_s H_{\rm ext}+K_2)} $, i.e., when $K_1$ is larger (smaller) than $K_2$, the principal of the ellipse is shorter (longer) in the $m_x$-direction than in the $m_y$-direction.  When the external field is large, $\mu_0 M_s H_{\rm ext} \gg K_1, K_2$, the precession becomes circular. Throughout this article we consider $K_1, K_2, H_{\rm ext}$ larger than zero so that the magnetic precession is stable. 

The linearized equations of motion follow from inserting the approximation for $\vm$  for small $m_x$ and $m_y$ into the Landau-Lifshitz-Gilbert Eq.~(\ref{eq:LLfirst}), which yields
\begin{subequations}
	\label{eq:linearizedmxmy}
\begin{eqnarray}
 -\dot m_x (t) &=& \frac{\gamma}{M_s} \frac{\partial E_0[\vm]}{\partial m_y} = \omega_ 2 m_y~, \\
 \dot m_y (t) &=& \frac{\gamma}{M_s} \frac{\partial E_0[\vm]}{\partial m_x}=\omega_1 m_x~,
\end{eqnarray}
\end{subequations}
with the frequencies $\omega_1 = \gamma \left( \mu_0 H_{\rm ext} + K_1/M_s \right)$ and $\omega_2 = \gamma \left( \mu_0 H_{\rm ext} + K_2/ M_s \right)$. These equations are most conveniently solved by rewriting them in terms of a radial coordinate $r (t)$ and an angle $\varphi (t)$  according to 
\begin{subequations}
	\label{eq:parametrizatiomxmy}
	\begin{eqnarray}
  m_x (t) &=& \sqrt{\frac{\sqrt{\omega_1 \omega_2}}{\omega_1} }r(t) \cos \varphi (t)~, \\
   m_y (t) &=& \sqrt{\frac{\sqrt{\omega_1 \omega_2}}{\omega_2}} r(t) \sin \varphi (t)~,
\end{eqnarray}
\end{subequations}
which parametrizes the elliptical motion in the $m_x-m_y$-plane, and where the overall factor $\left(\omega_1 \omega_2\right)^{1/4}$ is included to make $r(t)$ dimensionless. Inserting Eqs.~(\ref{eq:parametrizatiomxmy}) in the energy in Eq.~(\ref{eq:energylinearized}) yields
\begin{equation}
 E_0 [r] = \mu_0 M_s \frac{\sqrt{\omega_1 \omega_2}}{2} r^2~,
\end{equation}
whereas the equations of motion for $r(t)$ and $\varphi (t)$ are found by inserting the parameterization Eqs.~(\ref{eq:parametrizatiomxmy}) into Eqs.~(\ref{eq:linearizedmxmy}). These equations of motion are
\begin{subequations} \label{eq:eomactionangleellips}
	\begin{eqnarray}
 \dot r (t) &=& 0~,\\
 \dot \varphi (t)&=& \sqrt{\omega_1 \omega_2} = \frac{1}{\mu_0 M_s} \frac{\partial E_0 [r]}{\partial \left ( \frac{1}{2} r^2 \right)}~,
\end{eqnarray}\end{subequations}
which, afer comparison to Eqs.~(\ref{eq:actionangleeomgeneral}) shows that $\varphi$ is indeed an angle variable that is conjugate to the action variable $I \propto r^2$. The latter action variable corresponds, as expected, up to prefactors, to the area of the ellipse in the $m_x-m_y$-plane, and is proportional to the energy as well. From the above equations of motion it is found directly that the frequency of precession is $\sqrt{\omega_1 \omega_2}$.

Next we consider the case that $\psi$ is still time-independent --- but nonzero --- so that the elliptical trajectories (in the $m_x-m_y$-plane) are rotated as well (see Fig.~\ref{fig:illdynamics}), but with $\vn$ still pointing in the $z$-direction. Up to quadratic order in $m_x$ and $m_y$, the energy changes into
\begin{eqnarray}
\label{eq:energyrotatedlinearized}
E_\psi [\vm] &=& \frac{\mu_0 M_s H_{\rm ext}}{2} \left( m_x^2 +  m_y^2 \right) 
\nonumber \\
&&+ \frac{1}{2}\left(K_1  m_x^2+K_2  m_y^2\right) \cos^2 \psi \nonumber \\
&&+ \frac{1}{2}\left(K_2  m_x^2+K_1  m_y^2\right) \sin^2 \psi \nonumber \\
&&+ \frac{1}{2} \left( K_2 -K_1\right) m_x m_y \sin \left(2 \psi\right)~,
\end{eqnarray}
while the equations of motion become 
\begin{subequations}
	\label{eq:linearizedmxmyEpsi}
\begin{eqnarray}
-\dot m_x (t) &=& \frac{\gamma}{ M_s} \frac{\partial E_\psi[\vm]}{\partial m_y} ~, \\
\dot m_y (t) &=&\frac{\gamma}{ M_s} \frac{\partial E_\psi[\vm]}{\partial m_x}~.
\end{eqnarray}\end{subequations}
These latter equations of motion contain considerably more terms than the ones in Eqs.~(\ref{eq:linearizedmxmy}). Physically, the magnetization dynamics corresponds to the same elliptical precession as for $\psi=0$, but with the ellipse now rotated around the $z$-direction by an angle $\psi$. The equations of motions are therefore most straightforwardly solved by parameterizing $m_x$ and $m_y$ with the rotated version (over an angle $\psi$) of Eqs.~(\ref{eq:parametrizatiomxmy}), i.e., by
\begin{subequations}
	\label{eq:parametrizatiomxmypsi}
\begin{eqnarray}
\frac{m_x (t)}{\left(\omega_1 \omega_2\right)^{\frac{1}{4}}} &=&  r(t) \left[\cos \psi \frac{ \cos \varphi (t)}{\sqrt{\omega_1}} \right. \nonumber \\ && \left. + \sin \psi\frac{  \sin \varphi (t)}{\sqrt{\omega_2}}  \right]~,\\
\frac{m_y (t)}{\left(\omega_1 \omega_2\right)^{\frac{1}{4}}} &=&  r(t) \left[\cos \psi \frac{ \sin \varphi (t)}{\sqrt{\omega_2}} \right. \nonumber \\ && \left.- \sin \psi\frac{  \cos \varphi (t)}{\sqrt{\omega_1}}  \right].
\end{eqnarray} \end{subequations}
Inserting this parameterization into the energy $E_\psi$ and the equations of motions in Eqs.~(\ref{eq:linearizedmxmyEpsi}) yields  equations of motion for $r(t)$ and $\varphi(t)$ that are the same as Eqs.~(\ref{eq:eomactionangleellips}), as expected. In both the parameterization in Eqs.~(\ref{eq:parametrizatiomxmy}) and Eqs.~(\ref{eq:parametrizatiomxmypsi}) the angle variable $\varphi (t)$ corresponds to the angle between the vector $(m_x,m_y)$ and one of the principle axes of the ellipse. In terms of its relation to the fixed laboratory coordinates, its definition has, however changed. In the next section, we will see that this change in angle variable may lead to a geometric contribution to the angle variable for an adiabatically-slowly varying $\psi$.

\subsubsection{Time-dependent anisotropy}
\label{subsubsec:timedptellipse}

We now consider the situation that $\vn$ remains fixed to point in the $z$-direction, while $\psi$ is taken to be time-dependent, i.e., the anisotropy varies in time. In particular, we take $\psi = \psi_0 (t)$ to result in a loop in the space of parameters which determine the anisotropy, so with $\psi_0 (0) = \psi_0 (T)=0~({\rm mod}~2 \pi)$ and  
\begin{equation}
\psi_0 (t) = \frac{2 \pi t}{T}~{\rm for}~0 \leq t \leq T~,
\end{equation}
with $T$ the time over which $\psi$ changes. We consider the adiabatic limit, which in this particular case means that $\psi$ changes so slowly that $\sqrt{\omega_1 \omega_2} \gg 1/T$. Physically, this implies that there are many cycles of precession during the time when the anisotropy is varied. 

In the case of a time-dependent angle $\psi$, the equations of motion are still given by Eqs.~(\ref{eq:energyrotatedlinearized}) and (\ref{eq:linearizedmxmyEpsi}), with $\psi \to \psi_0 (t)$. It is again convenient to parameterize $m_x$ and $m_y$ by action and angle variables, using Eqs.~(\ref{eq:parametrizatiomxmypsi}) with $\psi \to \psi_0 (t)$: 
\begin{subequations}
	\label{eq:parametrizatiomxmypsitdpt}
\begin{eqnarray}
\frac{m_x (t)}{\left(\omega_1 \omega_2\right)^{\frac{1}{4}}} &=&  r(t)\left[\cos \psi_0 (t) \frac{ \cos \varphi  (t)}{\sqrt{\omega_1}} \right.\nonumber \\
 && \left.+ \sin \psi_0 (t)\frac{  \sin \varphi (t)}{\sqrt{\omega_2}}  \right]~, \\
\frac{m_y (t)}{\left(\omega_1 \omega_2\right)^{\frac{1}{4}}} &=&  r(t) \left[\cos \psi_0 (t) \frac{ \sin \varphi (t)}{\sqrt{\omega_2}} \right. \nonumber \\
&& \left. - \sin \psi_0 (t) \frac{ \cos \varphi (t)}{\sqrt{\omega_1}}  \right]~. 
\end{eqnarray}\end{subequations}
This means that at each time $t$, the instantaneous angle variable is indeed $\varphi$ because the instantaneous energy does not depend on it. Inserting the above parametrization in the equations of motion generates, however, extra terms as compared to Eqs.~(\ref{eq:eomactionangleellips}) because of the time-dependence of $\psi_0 (t)$ on which the time derivative acts. These extra terms give, ultimately, rise to the geometric contributions that we are after. In the first instance, we find, by inserting the above {\it ansatz} into Eqs.~(\ref{eq:energyrotatedlinearized}) and (\ref{eq:linearizedmxmyEpsi}) with $\psi \to \psi_0 (t)$, that
\begin{subequations}
	\label{eq:randthetatdptellipse1st} 
\begin{eqnarray}
\dot r (t)
 &=& \frac{\left(\omega_2-\omega_1\right)}{\sqrt{\omega_1 \omega_2}} r (t) \cos \varphi (t) \sin \varphi (t) \dot \psi_0 (t)~, \\
 \dot \varphi (t) &=& \sqrt{\omega_1 \omega_2} + \frac{1}{2}\left(\sqrt{\frac{\omega_1}{\omega_2}} + \sqrt{\frac{\omega_2}{\omega_1}} \right) \dot \psi_0 (t) \nonumber \\
 && + \frac{\left(\omega_2-\omega_1\right)}{2\sqrt{\omega_1 \omega_2}} 
 \cos \left(2 \varphi (t)\right) \dot \psi_0 (t)~.
\end{eqnarray} \end{subequations}
In the adiabatic limit, $\varphi (t)$ depends approximately linearly on time, so that $\sin \varphi (t)$ and $\cos \varphi (t)$ oscillate. Denoting the time average over such oscillations by $\langle \cdots \rangle$, we have that 
\begin{subequations}
	\label{eq:randthetatdptellipseaveraged} 
\begin{eqnarray}
\langle \dot r (t) \rangle
&=& 0~,  \\
\langle \dot \varphi (t) \rangle &=& \sqrt{\omega_1 \omega_2} + \frac{1}{2}\left(\sqrt{\frac{\omega_1}{\omega_2}} + \sqrt{\frac{\omega_2}{\omega_1}} \right) \dot \psi_0 (t)~.
\end{eqnarray} \end{subequations}
The second equation in this result is integrated from $0$ to $T$ to find the change in precession angle after the adiabatic change of the anisotropy parameters is performed and the anisotropy has returned to its initial configuration. We find that $\Delta \varphi \equiv \int_0^T \langle \dot \varphi (t) \rangle  dt = \Delta \varphi_{\rm dyn} + \Delta \varphi_{\rm geo}$, with 
\begin{subequations}
	\label{eq:geophasepsizero}
\begin{eqnarray}
 \Delta \varphi_{\rm dyn} &=& \sqrt{\omega_1 \omega_2} T~, \\
 \Delta \varphi_{\rm geo} &=& \frac{1}{2}\left(\sqrt{\frac{\omega_1}{\omega_2}} + \sqrt{\frac{\omega_2}{\omega_1}} \right)  \int_0^T \dot \psi_0 (t) dt~.
\end{eqnarray} \end{subequations}
The first of these contributions, i.e., $\Delta \varphi_{\rm dyn}$, is the usual dynamic contribution that is not geometric. The other contribution, $ \Delta \varphi_{\rm geo}$, is geometric in the sense that it does not depend on the time-dependence of the loop in parameter space along which the system is taken adiabatically, but only on the geometry of the loop. In this particular case, this means that $\Delta \varphi_{\rm geo}$ does not depend on the path $\psi_0 (t)$, but only on its end points, i.e., 
\begin{eqnarray}
\label{eq:finalresultgeophasestdptellipticity}
 \Delta \varphi_{\rm geo} &=& \frac{1}{2}\left(\sqrt{\frac{\omega_1}{\omega_2}} + \sqrt{\frac{\omega_2}{\omega_1}} \right)  \int_0^T \dot \psi_0 (t) dt \nonumber \\
 &&= \pi \left(\sqrt{\frac{\omega_1}{\omega_2}} + \sqrt{\frac{\omega_2}{\omega_1}} \right) ~,
\end{eqnarray}
where we used that $\psi_0(T)-\psi_0 (0)=2\pi$. The above result shows that when the anisotropy is varied adiabatically in such a away that the ellipse on which the magnetization precesses rotates $n$ times, the precession angle picks up a geometric contribution $ \pi  n\left(\sqrt{\omega_1/\omega_2} + \sqrt{\omega_2/\omega_1} \right) $. In case the precession is circular, i.e., when $\omega_1=\omega_2$, this angle is $2 \pi n$, and thus zero (mod $2\pi$). This result is a relation between the geometric angle and the ratio of the lengths of the principal axes of the ellipse on which the precession takes place. It is similar to the example of  elliptical particle-motion in phase space considered by Hannay \cite{Hannay_1985}, and provides the answer to the exercise that this article started out with in the introduction. 

At this point, we mention a subtlety that arises from the parameterization in Eqs.~(\ref{eq:parametrizatiomxmypsitdpt}). Namely, the energy $E_\psi [\vm]$ is invariant when $\psi \to \psi+ \pi$ whereas the parameterization in Eqs.~(\ref{eq:parametrizatiomxmypsitdpt}) is not. As a result, when $\psi_0 (t)$ is taken to vary adiabatically from e.g. $0$ to $\pi$, the energy returns to its value at $t=0$ but there appears to be a geometric contribution to the angle $\Delta \varphi_{\rm geo} = \pi \neq 0$ even when $\omega_1=\omega_2$.  This is, however, not a true geometric angle but rather a result of the parameterization not being invariant under $\psi_0 \to \psi_0+ \pi$. To isolate the true geometric contribution, paths that take $\psi_0 (t)$ from some value $\tilde \psi$ to $\tilde \psi + 2 \pi n$ should be considered, such that the parameterization in Eqs.~(\ref{eq:parametrizatiomxmypsitdpt}) ``makes a full loop" in the parameter space of $\psi_0$. The geometric angle in Eqs.~(\ref{eq:geophasepsizero}) should be computed using only such paths. To compute the geometric contribution to the angle that is acquired when  $\psi_0 (t)$ is taken to vary adiabatically from $\tilde \psi$ to $\tilde \psi+ \pi$, this result should be divided by two, {\it after} taking mod $2\pi$. 

Finally, note that we can always redefine the angle variable by adding a constant to it, i.e., by replacing $\varphi (t) \to \varphi (t) + \varphi_0$ with $\varphi_0$ independent of time in Eqs.~(\ref{eq:parametrizatiomxmypsitdpt}). This redefinition is similar to a gauge transformation, and leaves the geometric angle that the system picks up after it is taken along a close path in parameter space invariant.

\subsubsection{Circular precession in a time-dependent magnetic field}
\label{subsubsec:circulartdptfield}

Next, we consider the situation without ellipticity, i.e., $K_1=K_2\equiv K $, but take the direction of the external field to be arbitrary. Let $R(\theta, \phi)$ be the rotation matrix that rotates the $\vn$-direction to the direction of ${\bf e}_z$, i.e., $R (\theta, \phi) \vn = {\bf e}_z$, with ${\bf e}_z$ the unit vector in the $z$-direction. Then, by construction, inserting $ \vm (t)=R^{-1}  (\theta, \phi)\cdot (r(t) \sin \varphi(t),r(t) \cos \varphi (t),1)$ into the Landau-Lifshitz equation leads to the  equations of motion (\ref{eq:eomactionangleellips}) for small $r(t)$ (taking $K_1=K_2=K$) which shows that $\varphi (t)$ is an appropriate angle variable. Physically, this angle variable corresponds to the angle between the magnetization direction $\vm (t)$, projected on the plane perpendicular to $\vn$, and the ${\bf e}'_x$-direction.  The variable $r(t)$ is the radius of the --- in this case circular --- precession. 

We now take the direction of the external field to be time-dependent, i.e.,  $\vn (t) = (\sin \theta (t) \cos \phi (t), \sin \theta (t) \sin \phi (t), \cos \theta (t))$. Insertion of $ \vm (t)=R^{-1}   (\theta (t), \phi (t))\cdot (r(t) \sin \varphi(t),r(t) \cos \varphi (t),1)$ into the Landau-Lifshitz equation (\ref{eq:LLfirst}) yields in the first instance the equations of motion
\begin{subequations}
\begin{eqnarray}
 \dot r (t) &=& - \left[ \cos \varphi (t) \dot \theta (t) + \sin \varphi (t) \sin \theta (t) \dot \phi (t) \right], \\
 \dot \varphi (t) &=& \omega_{1}  + \frac{\sin \varphi (t) \dot \theta (t) + \cos \varphi (t) \sin \theta (t) \dot \phi (t)}{r(t)} \nonumber \\
 && - \cos \theta (t)  \dot \phi (t)~.
\end{eqnarray} \end{subequations}
The adiabatic limit physically corresponds to the case that the precession completes many cycles while the direction of the field changes slowly, so that  $|d \vn/dt| \ll \omega_1$ where $\omega_1=\omega_2=\gamma (\mu_0 H_{\rm ext} + K/ M_s)$ for the case that $K_1=K_2=K$. In this limit, we then average over the oscillating terms in the above equation which yields
\begin{subequations}
	\label{eq:actionangleaveragedtdptfield}
\begin{eqnarray}
\langle \dot r (t) \rangle &=& 0~, \\
\langle \dot \varphi (t) \rangle &=& \omega_{1}   - \cos \theta (t)  \dot \phi (t)~.
\end{eqnarray}\end{subequations}
Like before, we consider that the system is taken adiabatically along a loop in parameter space, i.e., we consider the direction $\vn$ of the field to make a loop on the surface of the unit sphere starting at time $t=0$ and ending at $t=T$. Integrating the second equation in (\ref{eq:actionangleaveragedtdptfield}) over time, we find that the first term gives a dynamic contribution $\Delta \varphi_{\rm dyn} = \omega_1 T$. The second term gives the geometric contribution
\begin{eqnarray}
\Delta \varphi_{\rm geo} &=& -\int_0^T \cos \theta (t) \dot \phi (t) \nonumber \\
&=&  \int_0^T \left[1-\cos \theta (t)\right] \dot \phi (t)
= \Omega~.
\end{eqnarray}
In going from the first to second line in the above, we added a multiple of $2\pi$, which is allowed because $\Delta \varphi_{\rm geo}$ is defined modulo $2\pi$. This allows us to rewrite the geometric contribution as the area $\Omega$ enclosed by the path $\vn (t)$ on the unit sphere. This result is understood as follows: the geometric contribution to the angle variable is the same as the angle over which a vector, that is transported parallel on the unit sphere, rotates,  which is well known to be $\Omega$. 

\subsubsection{Elliptical precession in a time-dependent magnetic field}
\label{subsubsec:ellipsetdptfield}
We now consider the geometric contribution to the precession angle that results from a time-dependent adiabatic excursion of both the ellipticity and the direction of the external field. To this end, we consider the energy of our toy model in Eq.~(\ref{eq:fullenergytoymodel}) in the most general case $K_1 \neq K_2$, and arbitrary and time-dependent direction of field $\vn (t)$, and direction of anisotropy as parametrized by  $\psi (t)$. The magnetization direction is now written in terms of action and angle variables $r (t)$ and $\varphi (t)$ by combining the transformation of the previous section with Eqs.~(\ref{eq:actionangleaveragedtdptfield}), i.e., by using $\vm (t) = R^{-1} (\theta(t), \phi (t)) \cdot (m_x (t), m_y (t), 1)$, with $m_x (t)$ and $m_y (t)$ given by  Eqs.~(\ref{eq:parametrizatiomxmypsitdpt}). Inserting this in the Landau-Lifshitz equation gives, after averaging over oscillatory terms, the equations of motion
\begin{subequations}\label{eq:actionangleaveragedtdptfieldandellipticity}
\begin{eqnarray}
\langle \dot r (t) \rangle &=& 0~,\\
\langle \dot \varphi (t) \rangle &=&  \sqrt{\omega_{1}\omega_2} + 
 \frac{1}{2}\left(\sqrt{\frac{\omega_1}{\omega_2}} + \sqrt{\frac{\omega_2}{\omega_1}} \right) \nonumber \\ && \times \left[\dot \psi_0 (t) - \cos \theta (t)  \dot \phi (t)\right]~.
\end{eqnarray} \end{subequations}
The geometric angle after a cyclic adiabatic excursion from $t=0$ to $t=T$ is found from this latter result as
\begin{eqnarray}
\label{eq:geophasetoymodelgeneral}
\Delta \varphi_{\rm geo} &=& 
\frac{1}{2}\left(\sqrt{\frac{\omega_1}{\omega_2}} + \sqrt{\frac{\omega_2}{\omega_1}} \right) \nonumber \\ && \times \int_0^T dt \left[\dot \psi_0 (t) - \cos \theta (t)  \dot \phi (t)\right]~,
\end{eqnarray}
which is the sum of a contribution due to time-dependent ellipticity and the time-dependent direction of the field. 
This result shows that the ellipticity of the precession affects the geometric angle resulting from adiabatically changing the direction of field, i.e., the second term in the above, making it impossible to express it in terms of the path enclosed by the area on the unit sphere. One way to understand this is as follows. The anisotropy breaks spin conservation and therefore leads to nutation. This makes it not straightforward to view the adiabatic dynamics as parallel transport. Note that the contribution due to adiabatic variation of the ellipticity, the first term in Eq.~(\ref{eq:geophasetoymodelgeneral}), is the same as found in Sec.~\ref{subsubsec:timedptellipse}. 

While, in principle, the geometric contribution to the precession angle could be measured directly, this may be very hard to do in practice because it would involve time-resolved measurements of small deviations of the magnetization. Often, one would rely on some form of interference set-up. To perform this interference in the time domain, however, may again be very hard because the precession relaxes on a time scale $1/\alpha \sqrt{\omega_1 \omega_2}$. In the next section, we therefore discuss a generalization of the geometric angles to the position domain and, in particular, the geometric angles that can be acquired by a spin wave.

\section{Spin waves}
\label{sec:spinwaves}

In this section, we consider a different context in which the Hannay angles discussed in the previous section may arise. Namely, we consider the propagation of a spin wave. Such a wave may, e.g., pick up a geometric angle when the parameters in the energy change as a function of position. This geometric angle may be used to manipulate the spin wave. Manipulation of spin waves is the goal of the field that is nowadays dubbed magnonics \cite{Kruglyak_2010}. From now on, we focus on the Hannay angle due to adiabatically-varying ellipticity. This particular geometric angle was first discussed in Refs.~\cite{PhysRevLett.93.247202, PhysRevB.72.024456}. In general, geometric phases for spin waves are often referred to as magnon Berry phases. 

To consider spin waves, the expression that we used for the energy until now needs to be modified to include exchange. Starting from the expression in Eq.~(\ref{eq:energylinearized}), we have that up to quadratic order
\begin{eqnarray}
\label{eq:energylinearizedplusexchange}
&& E_0^{\rm ex} [\vm] = \int d \bx \left[\frac{\mu_0 M_s H_{\rm ext}}{2} \left( m_x^2 +  m_y^2 \right) + \frac{K_1}{2}  m_x^2   \right. \nonumber \\
&& \left. +\frac{K_2}{2}  m_y^2 -\frac{J_s}{2} \left(m_x \nabla^2 m_x + m_y \nabla^2 m_y \right) \right]~,
\end{eqnarray}
where the deviations $m_x$ and $m_y$ are now a function of both time $t$ and position $\bx$, and $J_s$ is the exchange stiffness. The linearized equations of motion follow analogously to Eqs.~(\ref{eq:linearizedmxmy}) and are given by
\begin{eqnarray}
\label{eq:linearizedmxexchange}
&&-\dot m_x (\bx, t) = \frac{\gamma}{ M_s} \frac{\delta E^{\rm xc}_0[\vm]}{\delta m_y} \nonumber \\
&&= \gamma \left[ \mu_0 H_{\rm ext}  + \frac{1}{ M_s} \left(-J_s \nabla^2 + K_2\right)\right] m_y~,  \end{eqnarray}
and
\begin{eqnarray}
\label{eq:linearizedmyexchange}
&&\dot m_y (\bx, t) = \frac{\gamma}{ M_s} \frac{\delta E^{\rm xc}_0[\vm]}{\delta m_x} \nonumber
 \\
 &&= \gamma \left[ \mu_0 H_{\rm ext}  + \frac{1}{ M_s} \left(-J_s \nabla^2 + K_1\right)\right] m_x~.
\end{eqnarray}
Spin waves correspond to plane-wave solutions of these latter two equations. In keeping with our discussion in terms of Hannay angles of the previous section, we write these plane wave solutions in terms of action and angle variables $r (t)$ and $\varphi (t)$ that are now defined as 
\begin{subequations}\label{eq:parametrizatiomxmyexchange}
	\begin{eqnarray}
m_x (\bx, t)\!\!\!&=&\!\!\!\sqrt{\frac{\sqrt{\omega_1 (k) \omega_2 (k)}}{\omega_1 (k)} } r(t) \cos \left[ \varphi (t)\!- \!\bk \cdot \bx \right], \\
m_y (\bx,t)\!\!\!&=&\!\!\!\sqrt{\frac{\sqrt{\omega_1 (k) \omega_2 (k)}}{\omega_2 (k)}} r(t) \sin\left[\varphi (t)\!-\!\bk \cdot \bx\right],
\end{eqnarray} \end{subequations}
in which the frequencies  $\omega_1 (k) = \gamma \left(\mu_0 H_{\rm ext} + K_1/ M_s +J_s k^2/ M_s \right)$  and $\omega_2 (k) = \gamma \left( \mu_0 H_{\rm ext} + K_2/ M_s +J_s k^2/ M_s \right)$ now incorporate exchange, and where $\bk$ is the wave vector of the spin wave. The resulting equations are $\dot r (t)=0$, as expected, and $\dot \varphi (t)=\sqrt{\omega_1 (k) \omega_2 (k)} \equiv \omega_{\rm sw}(k)$ which gives the spin-wave dispersion $\omega_{\rm sw} (k)$. 

\begin{figure} 
	\includegraphics[width=9cm]{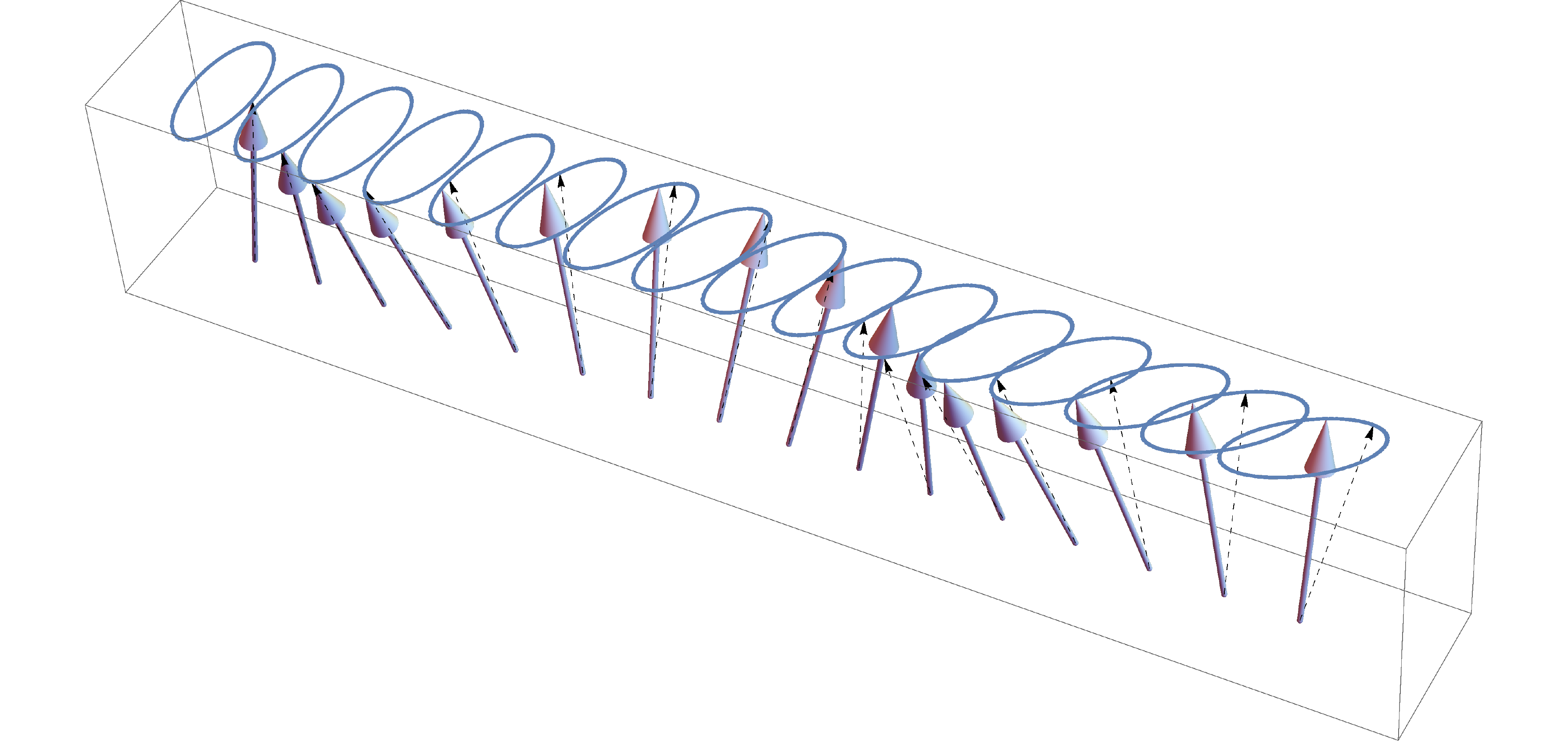} 
	\caption{Illustration of a spin wave that travels through a region in which the anisotropy axes vary with position. The ellipses correspond to the ellipse of precession that is favoured by the local anisotropy. The solid arrows illustrate a snapshot of a spin wave that has acquired the geometric phase shift. The phase shift is illustrated by comparing to the spin wave without geometric phase shift, as indicated by the dashed arrows.}
	\label{fig:spinwave}
\end{figure}

The spin-wave solutions in Eqs.~(\ref{eq:parametrizatiomxmyexchange}) correspond to elliptically-precessing spin waves. If the anisotropy varies in space, the direction of the principal axes of the ellipse of this precession will also vary in space. To explore how this gives rise to geometric angles we consider the model of Eq.~(\ref{eq:energyrotatedlinearized}) and generalize it to the case of position-dependent anisotropy $\psi=\psi (x)$. For simplicity we take the anisotropy to vary in the $x$-direction only, and will take the spin wave to propagate in this direction as well.  The energy is then given by 
\begin{eqnarray}
\label{eq:energyrotatedlinearizedexc}
E^{\rm ex}_\psi [\vm] &=& \int d \bx \left[ \frac{\mu_0 M_s H_{\rm ext}}{2} \left( m_x^2 +  m_y^2 \right)\right. 
\nonumber \\
&&+ \frac{1}{2}\left(K_1  m_x^2+K_2  m_y^2\right) \cos^2 \psi \nonumber \\
&&+ \frac{1}{2}\left(K_2  m_x^2+K_1  m_y^2\right) \sin^2 \psi \nonumber \\
&& + \frac{1}{2} \left( K_2 -K_1\right) m_x m_y \sin \left(2 \psi\right) \nonumber \\
&& \left. -\frac{J_s}{2} \left(m_x \nabla^2 m_x + m_y \nabla^2 m_y \right) \right]~,
\end{eqnarray}
which yields the equations of motion
\begin{subequations}\label{eq:eomlinearizedfunctional}
\begin{eqnarray}
&&-\dot m_x (\bx, t) = \frac{\gamma}{ M_s} \frac{\delta E^{\rm xc}_\psi[\vm]}{\delta m_y}~, \\
&&\dot m_y (\bx, t) = \frac{\gamma}{ M_s} \frac{\delta E^{\rm xc}_\psi[\vm]}{\delta m_x}~.
\end{eqnarray}\end{subequations}
that contain considerably more terms than Eqs.~(\ref{eq:linearizedmxexchange})~and~Eqs.~(\ref{eq:linearizedmyexchange}) and are not written out explicitly. To accomodate for the position-dependent anisotropy and resulting ellipticity, we attempt solutions of the form 
\begin{eqnarray}
\label{eq:parametrizatiomxpsiexc}
&&\frac{m_x (\bx, t)}{\left(\omega_1 (k) \omega_2 (k)\right)^{\frac{1}{4}}} =  r(t) \left[\cos \psi (x) \frac{ \cos \left[ \omega  t  +\varphi (x)\right] }{\sqrt{\omega_1 (k)}} \right. \nonumber \\
&& \left. + \sin \psi (x) \frac{  \sin \left[ \omega  t  +\varphi (x)\right]}{\sqrt{\omega_2 (k)}}  \right]~,
\end{eqnarray}
and
\begin{eqnarray}
\label{eq:parametrizatiomypsiexc}
&& \frac{m_y (\bx, t)}{\left(\omega_1 (k) \omega_2 (k) \right)^{\frac{1}{4}}} =  r(t) \left[\cos \psi (x) \frac{ \sin \left[ \omega t +\varphi (x)\right]}{\sqrt{\omega_2 (k)}} \right. \nonumber \\ 
&&\left. - \sin \psi (x)\frac{  \cos \left[ \omega  t  +\varphi (x)\right]}{\sqrt{\omega_1 (k)}}  \right], 
\end{eqnarray}
which generalize Eqs.~(\ref{eq:parametrizatiomxmypsi}) to incorporate exchange and a position-dependent anisotropy, because the ellipse of precession is locally rotated over the angle $\psi (x)$ to the ellipse favored by the anisotropy. In Fig.~\ref{fig:spinwave} these spin waves are illustrated. The ellipses in this figure indicate the precession that is favoured by the local anisotropy. The solid arrows illustrate the spin wave with geometric phase shift. The dashed arrows correspond to the spin wave without the geometric phase.

Since the position-dependent anisotropy breaks translation invariance, the trial solution is labeled by a frequency $\omega$. The wave number $k$ is still to be determined and should be interpreted as a function of this frequency. We expect that these trial solutions are valid in the adiabatic limit when $|\psi'(x)| \ll k$, where the prime indicates a derivative with respect to $x$. Note that $\psi (x)$ is assumed to be a given function that is determined by how the anisotropy varies in space. 

Using the above trial solutions, $\varphi (x)$ is computed by inserting them into the equations of motion (\ref{eq:eomlinearizedfunctional}). This yields, after averaging as before over oscillations in time, that  $\langle \dot r (t) \rangle =0$ to lowest order in $\varphi'(x)$ and $\psi '(x)$. We define $\langle \varphi'(x)\rangle=-k+\delta \varphi $, where $\delta \varphi $ is the lowest nonzero order in $\psi'(x)$. To zeroth order in $\psi'(x)$ we find that the possible values of $k$ are determined by solving for $k$ in the equation $\omega = \omega_{\rm sw} (k)$. One of these wave vectors is imaginary and corresponds to an evanescent wave. We consider only the propagating wave, and call its wave vector $\kappa (\omega)$. [For an explicit expression, see Eq.~(\ref{eq:dispersionwithdamping}) below, and use that $\kappa = k_+$ when $\alpha=0$.] We find to lowest order in $\psi' (x)$ that 
\begin{equation}
\label{eq:geophase1spinwave}
\langle \varphi '(x) \rangle = -\kappa (\omega)+  
\left(  \frac{\frac{\omega}{\gamma}}{
	\sqrt{\frac{(K_1-K_2)^2}{4 M_s^2}+\frac{\omega^2}{\gamma^2}}}
\right)\psi'(x)~.
\end{equation}
The first term in the above then gives the usual phase $\kappa (\omega) x$ of a wave, whereas the second term is the geometric contribution. From this result it is clear that a constant can be added to the phase, and that this does not affect the geometric contribution. 

Using the above result, we find that a spin wave that travels from $x=x_i$ to $x=x_f$ through a region in which the direction of anisotropy changes adiabatically slowly in space, which is within our model parameterized by $\psi (x)$, acquires a geometric angle, or phase difference
\begin{equation}
\label{eq:geophase2spinwave}
\Delta \varphi_{\rm geo} = \left(
\frac{2}{\sqrt{\frac{\omega_1 (\kappa)}{\omega_2 (\kappa)}}
	+\sqrt{\frac{\omega_2 (\kappa)}{\omega_1 (\kappa)}}}
\right)\int_{x_i}^{x_f} \psi'(x) dx~,
\end{equation}
where we rewrote the prefactor of the integral in a different form to connect to the result in Eq.~(\ref{eq:finalresultgeophasestdptellipticity}), and where it should be kept in mind that $\kappa=\kappa (\omega)$.
Interestingly, the prefactor of the integral in the above result is the reciprocal of the prefactor in Eq.~(\ref{eq:finalresultgeophasestdptellipticity}). This difference between the cases of time-dependent and position-dependent anisotropy is attributed to the relative minus sign between the temporal and spatial derivates in one of the equations of motion [see  Eqs.~(\ref{eq:linearizedmxexchange})~and~(\ref{eq:linearizedmyexchange})].

The geometric phase difference considered here can be used to manipulate the spin waves, e.g. in devices which exploit interference. Within the Gilbert damping phenomenology, the length scale above which interference between spin waves is washed out is proportional to $1/\alpha$. More precisely, it is on the order of $\left(\partial \omega_{\rm sw}/\partial k \right) \left(1/\alpha \omega_{\rm sw}\right)$, where the first factor is the group velocity of the spin waves and the second factor their lifetime. The maximum destructive interference is reached when $\Delta \varphi_{\rm geo} = \pi ({\rm mod} 2 \pi)$. Taking for example $\int_{x_i}^{x_f} \psi'(x) dx=2\pi$, this situation is achieved when $\sqrt{\omega_1 (\kappa)/\omega_2 (\kappa)}$ is equal to $2 \pm \sqrt{3}$. Depending on the energy $\omega$ of the spin waves, this puts a condition on the anisotropy. Before we consider in more detail a device that illustrates this geometric phase, we discuss the relation between the Hannay angle due to ellipticity of the  precession and a geometric phase that occurs in the context of superfluidity and Bose-Einstein condensation.

\section{Intermezzo:  geometric phases in superfluids}
\label{sec:connections}
Using the discussed formalism, one can give a simple derivation and interpretation for geometric phases that are acquired by excitations that propagate on top of a flowing superfluid \cite{PhysRevLett.97.040401}. As we shall see, the superfluid density gives rise to --- using the language of magnetism --- nonzero ellipticity. Using the language that is more common for superfluidity and superconductivity, nonzero ellipticity corresponds to nonzero anomalous averages of field operators. This couples particles and holes and requires one to perform a Bogoliubov transformation to a new basis to find the proper excitations. The resulting Bogoliubov quasiparticles may then pick up geometric phases in case the phase of the superfluid order parameter is time-dependent or position-dependent. We consider for simplicity only the latter case here as it maps one-to-one to the problem treated in Sec.~\ref{sec:spinwaves}. An interesting generalization would be to consider a time-dependent spinor superfluid, as this situation could be mapped to the cases of Sec.~\ref{sec:singledomain}.

The simplest description of a homogeneous superfluid of particles with mass $m$ is the Gross-Pitaevskii equation for the superfluid order parameter $\Psi (\bx,t)$ given by \cite{pítajevskíj2003bose}
\begin{equation}
\label{eq:gpequation}
 i \hbar \frac{\partial \Psi (\bx,t)}{\partial t} = \left[ -\frac{\hbar^2 \nabla^2}{2m} - \mu +g \left|\Psi (\bx,t)\right|^2 \right] \Psi (\bx,t)~,
\end{equation}
with $\hbar$ the reduced Planck's constant, $\mu$ the chemical potential, and $g>0$ a parameter that governs the strength of the interactions between the particles. For a time-independent flowing superfluid we have that $\Psi_0 (\bx) = \sqrt{n} e^{i \vartheta (\bx)}$, where $n=-\mu/g$ is the superfluid density, and that the superfluid velocity ${\bf v}_s (\bx)$ is related to the phase via ${\bf v}_s (\bx)=\hbar \nabla  \vartheta (\bx)/m$. We linearize the Gross-Pitaevskii equation around this time-independent situation via $\Psi (\bx,t) = \Psi_0 (\bx) + \delta \Psi (\bx,t)$, which yields the Bogoliubov-de Gennes equations
\begin{eqnarray}
\label{e:bogodegennes} 
&& i \hbar \frac{\partial }{\partial t} \left(
 \begin{array}{c}
\delta \Psi (\bx,t) \\ -\delta \Psi^* (\bx,t)
 \end{array}
 \right) = \nonumber \\
&& \left(
 \begin{array}{cc}
 -\frac{\hbar^2 \nabla^2}{2m} + g n & g n e^{2 i \vartheta (\bx)}\\
g n e^{-2 i \vartheta (\bx)} &  -\frac{\hbar^2 \nabla^2}{2m} +gn
 \end{array}
 \right)   \! \cdot \! \left(
  \begin{array}{c}
 \delta \Psi (\bx,t) \\ \delta \Psi^* (\bx,t)
 \end{array}
 \right)\!\!,
\end{eqnarray}
for fluctuations on top of the superfluid. The Bogoliubov-de Gennes equations are equivalent to a special case of Eqs.~(\ref{eq:energyrotatedlinearizedexc})~and~(\ref{eq:eomlinearizedfunctional}), as is found by substituting the linearized Holstein-Primakoff transformation \cite{PhysRev.58.1098}
\begin{subequations}
\label{eq:linearizedHPtrafo}
\begin{eqnarray}
m_x &=& \frac{\Psi+\Psi^*}{2\sqrt{n}}~,\\
m_y &=& \frac{\Psi^*-\Psi}{2i\sqrt{n}}~,
\end{eqnarray} \end{subequations}
and making the replacements $\gamma/M_s \to 1/2 \hbar n$, $J_s \to \hbar^2 n/m$, $K_1 \to 4 g n^2$ and taking $H_{\rm ext}=K_2 =0$. We then find that the frequencies become $\omega_2 (k) = \hbar k^2/2m\equiv \epsilon (k)/\hbar$ and $\hbar \omega_1 (k) = \epsilon (k)+2gn$ that yields the famous Bogoliubov dispersion relation $E (k) = \hbar \sqrt{\omega_1 (k)\omega_2 (k)} = \sqrt{\epsilon (k) (\epsilon (k) + 2gn )} $ that is gapless and linear at long wavelengths. 

Using the results in Eqs.~(\ref{eq:geophase1spinwave})~and~(\ref{eq:geophase2spinwave}) we immediately find that a Bogoliubov quasi-particle with energy $\hbar \omega$ that propagates on top of a superfluid that flows with velocity $v_s (\bx)$ in the $x$-direction acquires the geometric phase 
\begin{eqnarray}
\label{eq:bogoliubovberryphase}
  \Delta \varphi_{\rm geo} = \frac{m}{\hbar}\left(\frac{\hbar \omega}{\sqrt{(g n)^2 + (\hbar\omega)^2}}\right) \int dx v_s (\bx)~.
\end{eqnarray}
Note that this geometric phase vanishes (mod $2\pi$) for large energies $\hbar \omega \gg g n$, and energies $\hbar \omega \to 0$, and will be most pronounced for energies $\hbar \omega \sim g n$. The prefactor in our result Eq.~(\ref{eq:bogoliubovberryphase}) is different from the prefactor of Ref.~\cite{PhysRevLett.97.040401}. While a direct comparison between our approach and the formalism of this work is hard, it is probably because in  Ref.~\cite{PhysRevLett.97.040401} a wave packet rather than a single wave is considered.

\section{Magnon transport}
\label{sec:magnons}

\begin{figure} 
\includegraphics[width=.7\columnwidth]{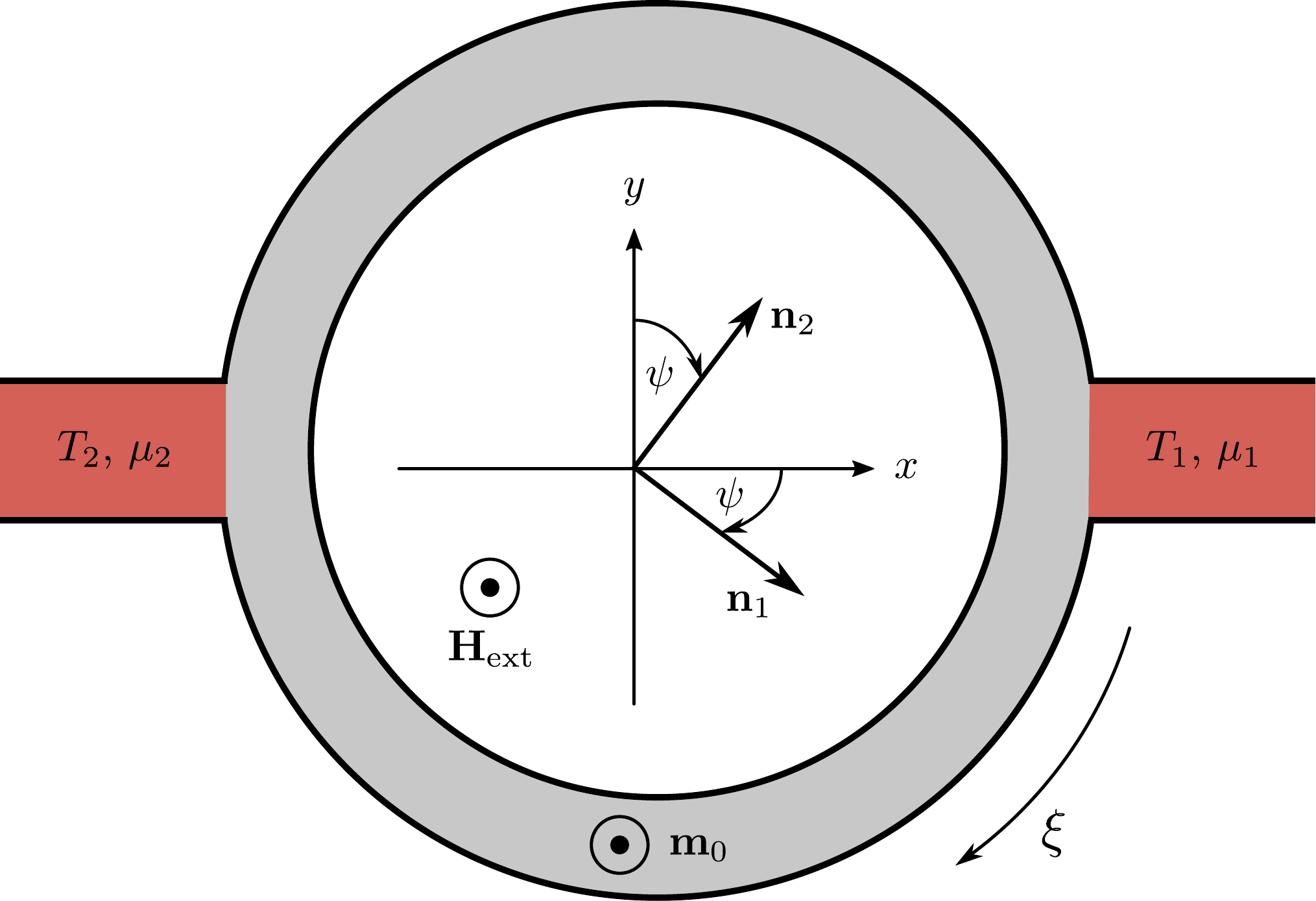} 
\caption{Sketch of the model we consider in Sec.~\ref{sec:magnons}: 
A ferromagnetic ring (grey) with two metallic leads (red)
attached on opposite sides.
The leads are kept at different temperatures $T_1$ and $T_2$
and may also have spin accumulations $\mu_1$ and $\mu_2$.
The magnetization ${\bf m}_0$ is aligned parallel to an 
external magnetic field ${\bf H}_{\rm ext}$ pointing out of the ring plane.
Magnons moving along the ring in $\xi$-direction accumulate a geometric phase
because the directions ${\bf n}_1$ and ${\bf n}_2$ of the in-plane
anisotropies $K_1$ and $K_2$ vary along the ring.
This change is parametrized by the angle $\psi$ that changes by $2\pi$
when moving around the ring once. 
}
\label{fig:ring}
\end{figure}

We have considered the Hannay angle acquired by single coherent spin waves due to position-dependent anisotropy, resulting in position-dependent ellipticity, in Sec.~\ref{sec:spinwaves}. For a thermal, and, therefore, incoherent distribution of spin waves this geometric angle will also have consequences. In the incoherent case we find it more appropriate to refer to linear excitations of the magnetic order as magnons rather than spin waves. To illustrate the effects of the geometric phase on incoherent magnon transport,
we consider a mesoscopic electrically-insulating magnetic ring of circumference $L$ as depicted in Fig.~\ref{fig:ring}.
In this system, the magnetization is aligned parallel to a static external magnetic field normal to the ring plane,
while the direction of the in-plane anisotropies, characterized by the same angle $\psi$ as before, slowly varies
from $\psi(0)=0$ to $\psi(L) = 2\pi$ along the ring, 
with $\psi\left(\frac{L}{2}\right) = \pi$. Here, the position coordinate on the ring is denoted as $\xi$ so that $\psi=\psi(\xi)$. (Throughout this section, we use $\psi'(\xi) \equiv d\psi (\xi)/d\xi$.)
Because of this change in the direction of the anisotropy axes, 
and hence of the principal axes of the elliptical spin-wave precession,
the magnons will accumulate a geometric phase when moving along the ring. We expect that this phase will give rise to interference effects that affect the magnon spin transport. As mentioned, such interference may be washed out by relaxation. To account in the simplest manner for such relaxation we include Gilbert damping in this section. 

At $\xi=0$ and $\xi=\frac{L}{2}$, there are metallic leads attached to the ring that 
enable electrical injection and detection of the spin current in the ring. Such non-local electrical injection and detection was developed by Cornelissen {\it et al.}, who used Pt contacts on top of the magnetic insulator yttrium-iron garnet \cite{ab9817b8a4964a7ebb40e9945e0961de}.
For simplicity, 
we further assume that the ring is narrow enough so that the magnons are essentially confined to one-dimensional propagation 
along the circumference of the ring,
and long enough that we can ignore the curvature of the ring.
Therefore we consider the Hamiltonian
\begin{eqnarray}
\label{eq:H_rotated}
{\cal H}^{\rm ex}_\psi 
&=& \int d \xi \biggl[ -\frac{J_s}{2s^2} {\bf s} \cdot \partial_\xi^2 {\bf s} - \hbar\gamma \mu_0 H_{\rm ext} s_z 
\nonumber \\
&&+ \frac{1}{2s^2}\left(K_1  s_x^2+K_2  s_y^2\right) \cos^2 \psi \nonumber \\
&&+ \frac{1}{2s^2}\left(K_2  s_x^2+K_1  s_y^2\right) \sin^2 \psi \nonumber \\
&& + \frac{1}{4s^2} \left( K_2 -K_1\right) \left( s_x s_y + s_y s_x \right) \sin \left(2 \psi\right) \biggr]~, \nonumber \\
&& 
\end{eqnarray}
which is the quantum-mechanical generalization of the classical energy (\ref{eq:energyrotatedlinearizedexc})
that describes spin waves in the presence of spatially-varying anisotropy axes. Furthermore, 
${\bf s}$ is the local spin density operator; 
it is related to the classical magnetization direction used in the preceding sections via ${\bf m}=\langle{\bf s}\rangle/s$,
where $s = M_s / \hbar \gamma$.
Magnons are introduced via a linearized Holstein-Primakoff transformation (see e.g. \cite{PhysRev.58.1098}):
\begin{subequations} \label{eq:HP_transformation}
\begin{eqnarray}
s_+ &=& s_-^\dagger = \sqrt{2s} \left[ a + {\cal O}(s^{-1}) \right] , \\
s_z &=& s - a^\dagger a ,
\end{eqnarray}
\end{subequations}
where
$s_\pm = s_x \pm i s_y$,
and the magnon creation operators $a^\dagger(x)$ satisfy the bosonic commutation relation
$[a(\xi),a^\dagger(\xi')] = \delta(\xi-\xi')$.
In terms of these magnon operators, 
the Hamiltonian (\ref{eq:H_rotated}) becomes
\begin{eqnarray}
\label{eq:H_boson}
{\cal H}^{\rm ex}_\psi 
&=& \hbar\gamma \int d \xi \biggl[  
a^\dagger \left( \mu_0 H_{\rm ext} + \frac{ K_1 + K_2 }{ 2 M_s } - \frac{ J_s }{ M_s} \partial_\xi^2 \right) a \nonumber\\
&&
+ \frac{ K_1 - K_2 }{ 4 M_s } \left( a^2 e^{2i\psi} + (a^\dagger)^2 e^{-2i\psi} \right)
\biggr] ,
\end{eqnarray}
where we dropped a constant contribution corresponding to the classical ground-state energy. Two remarks are now in order. First, from the above result one can explicity see that only when the magnons are elliptical, i.e., when $K_1-K_2 \neq 0$, there are anomalous terms [$\sim a^2$ and $~(a^\dagger)^2$] in the above Hamiltonian. These anomalous terms give rise to the nonzero anomalous averages that were already mentioned in Sec.~\ref{sec:connections}. The anomalous Green's functions that are introduced below are examples of such anomalous averages. The second remark is that we have adopted a quantum-mechanical approach. We find this convenient because it allows us to develop a theory for the magnon spin transport based on the non-equilibium Green's function formalism \cite{rammer2004quantum}, that straightforwardly incorporates the incoherent magnon distribution. This approach is, however, in the linear regime equivalent \cite{PhysRevB.96.174422} to a stochastic generalization of the Landau-Lifschitz-Gilbert equation that incorporates the incoherent magnon distribution via noisy magnetic fields. The geometric phases that are considered are therefore still classical. This is understood as they are, within the current formulation, phases of the magnon creation and annihilation operators that create respectively destroy excitations on top of the magnetically-ordered classical groundstate. These phases are therefore directly observable. For example, the complex phase of $\langle a \rangle$ determines the angle of the magnetization in the $x-y$-plane. 

\begin{figure} 
\includegraphics[width=.9\columnwidth]{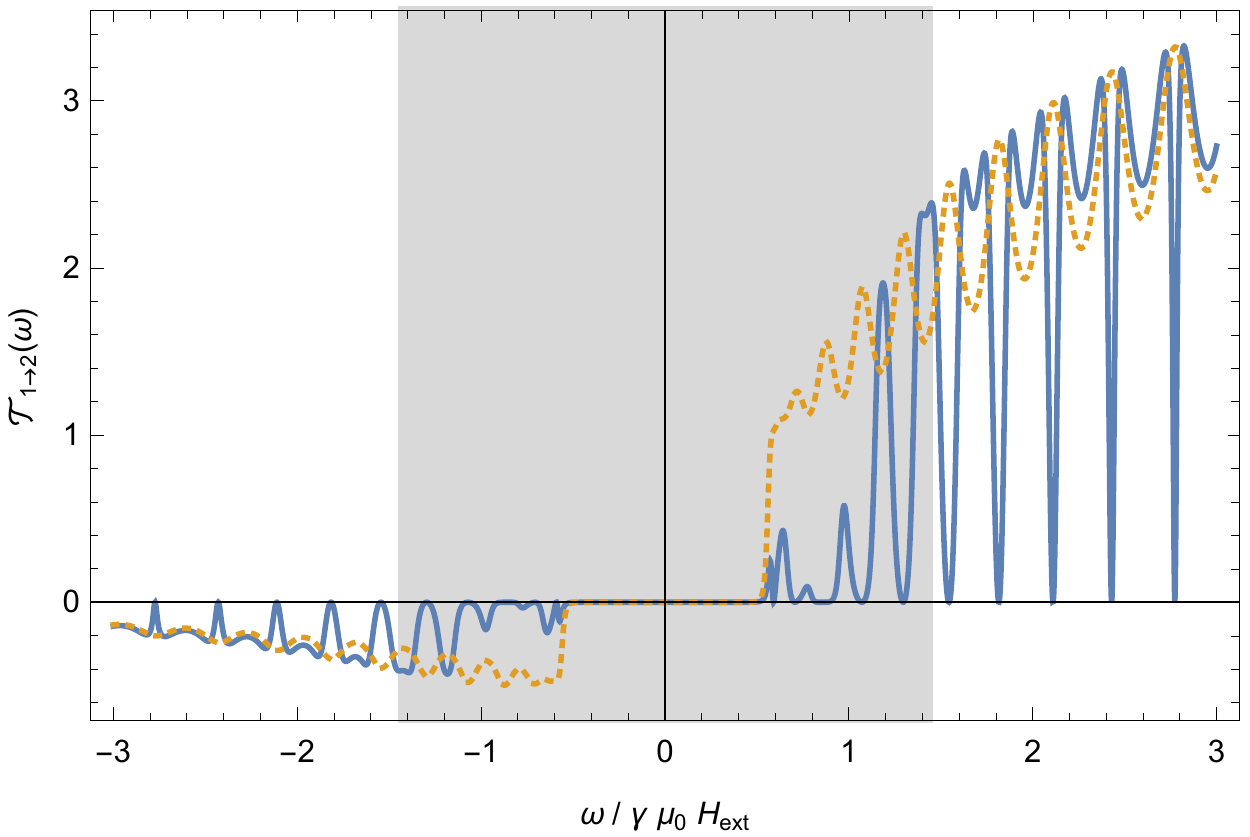} 
\caption{Transmission function (\ref{eq:transmission}) as function of frequency $\omega$ 
with (solid line) and without (dashed line) geometric phase.
The shaded area denotes the region 
$ |\omega| \le \gamma \left| K_1 - K_2 \right| / 2 M_s $
of destructive interference.
External magnetic field and anisotropy constants are 
$ \mu_0 H_{\rm ext} = 0.1\,{\rm T} $,
$ K_1 / M_s = 0.2\, {\rm T} $, and $ K_2 / M_s = -0.09\, {\rm T} $,
and the circumference of the ring is $L= 1.07 \, \mu{\rm m} $.
The interface couplings are set to
$ \alpha_1^{\rm sp} = \alpha_2^{\rm sp} = \alpha^{\rm sp} = 10.7 \,\textrm{nm} $,
and the lead spin accumulations are set to zero, 
$ \mu_1 = \mu_2 = 0 $.  
For the remaining material parameters, 
values of yttrium-iron garnet are used \cite{Cherepanov1993}: 
$ \hbar \gamma J_s / M_s = 8.458 \times 10^{-40} \, {\rm J}\,{\rm m}^2 $,
$ \hbar \gamma = 2 \mu_B $ (where $\mu_B$ is the Bohr magneton), and
$ \alpha = 10^{-4} $.
}
\label{fig:transmission}
\end{figure}

The spin current transmitted from the lead at $\xi=0$ to the lead at $\xi=\frac{L}{2}$
in a stationary state can be calculated with
the non-equilibrium Green's function technique \cite{rammer2004quantum};
the details of the computation are relegated to the Appendix.
The final result is 
\begin{eqnarray} \label{eq:current}
I_{1\to 2} 
&=& \int_{-\infty}^\infty \frac{ d \omega} {2\pi} {\cal T}_{1\to 2}(\omega) \nonumber\\
&& \times
\left[ f_B\left( \frac{\hbar \omega - \mu_2}{ k_B T_2 } \right) - f_B\left( \frac{\hbar \omega - \mu_1}{ k_B T_1 } \right) \right] ,
\nonumber\\
\end{eqnarray}
where
$f_B(x) = 1/(e^x-1)$
is the Bose function,
$T_{1/2}$ and $\mu_{1/2}$ are the temperature and spin accumulation in each lead,
and the transmission function is given by
\begin{eqnarray} \label{eq:transmission}
{\cal T}_{1\to 2}(\omega) 
&=& 2 \alpha_1^{\rm sp} \alpha_2^{\rm sp} \left( \hbar\omega - \mu_1 \right)
\nonumber\\
&& \times \Biggl[
\left( \hbar\omega - \mu_2 \right)
\left| g^R\left( \frac{L}{2}, 0; \omega \right) \right|^2 
\nonumber\\
&&
\phantom{ \times \Biggl[ }
- \left( \hbar\omega + \mu_2 \right)
\left| \tilde{g}^R\left( \frac{L}{2}, 0; \omega \right) \right|^2
\Biggr] .
\nonumber\\
\end{eqnarray}
Here,
$\alpha_{1/2}^{\rm sp}$ characterizes the interfacial coupling of magnons and lead electrons and 
is proportional to the spin-mixing conductance \cite{PhysRevLett.84.2481}. When the spin accumulation in the leads is zero, the interfacial coupling gives rise to an enhancement, localized at the interface,  of the Gilbert damping of the homogeneous mode. This enhancement is characterized by $\alpha_{1/2}^{\rm sp}$. 
Furthermore, $g^R\left( \frac{L}{2}, 0; \omega \right)$
and $\tilde{g}^R\left( \frac{L}{2}, 0; \omega \right)$
are the Fourier transforms of the normal and anomalous retarded magnon Green's functions
that describe the propagation of magnons from the lead at $\xi=0$ to the lead at $\xi=\frac{L}{2}$. 
Taking into account both the interfacial lead couplings and bulk Gilbert damping $\alpha$, 
the retarded Green's functions satisfy the Bogoliubov-de Gennes equations in frequency space that are explicitly stated in the Appendix. In the adiabatic limit of slowly varying $\psi(\xi)$, 
a solution of these Bogoliubov-de Gennes equations (\ref{eq:BdG_ring}) is obtained with the ansatz
\begin{subequations}
\begin{eqnarray}
g^R(\xi,0;\omega) 
&=& u(\omega) e^{-i\psi(\xi)+i\varphi(\xi,\omega)} , 
\\
\tilde{g}^R(\xi,0;\omega) 
&=& v(\omega) e^{i\psi(\xi)+i\varphi(\xi,\omega)} , 
\end{eqnarray}
\end{subequations}
where 
$\varphi(\xi,\omega) = k(\omega)\xi + \delta \varphi(\omega) \psi(\xi)$,
with $\delta \varphi$ denoting the coefficient of the geometric phase of the magnon.
To zeroth order in $\psi'$, one finds 
the dispersion relation of damped, elliptical spin waves:
\begin{eqnarray} 
&&
(1+\alpha^2) \omega^2 + 2i \alpha \gamma \omega 
\left( \mu_0 H_{\rm ext} + \frac{ K_1 + K_2 }{ 2 M_s } + \frac{J_s}{M_s} k^2 \right) 
\nonumber\\
&=& \omega_{\rm sw}^2(k),
\label{eq:omega_sw}
\end{eqnarray}
where 
$\omega_{\rm sw}(k) = \sqrt{\omega_1 (k) \omega_2 (k)} $,
with
$\omega_1 (k) = \gamma \left( \mu_0 H_{\rm ext} + K_1/ M_s +J_s k^2/ M_s \right)$  
and 
$\omega_2 (k) = \gamma \left( \mu_0 H_{\rm ext} + K_2/ M_s +J_s k^2/ M_s \right)$
as found previously in Sec.~\ref{sec:spinwaves}.
The dispersion equation (\ref{eq:omega_sw}) may be solved to obtain the possible magnon wave-vectors:
\begin{eqnarray}
\label{eq:dispersionwithdamping}
\frac{J_s}{M_s} k_{\pm}^2(\omega) 
&=& 
\pm \sqrt{ \frac{ \left( K_1 - K_2 \right)^2 }{ 4 M_s^2 } + \frac{\omega^2}{\gamma^2} } 
\nonumber\\
&&
- \mu_0 H_{\rm ext} - \frac{ K_1 + K_2 }{ 2 M_s } + i \alpha \frac{\omega}{\gamma} .
\end{eqnarray}
Note that only the $k_+$ solutions are traveling spin waves,
whereas the $k_-$ solutions are evanescent.
To first order in $\psi'$,
one finds 
\begin{equation}
\delta\varphi_\pm(\omega) = \pm 
\frac{ \frac{\omega}{\gamma} }{ \sqrt{ \frac{ \left( K_1 - K_2 \right)^2 }{ 4 M_s^2 } + \frac{\omega^2}{\gamma^2} }  } 
\end{equation}
for the additional geometric correction to the magnon phase,
which is again in agreement with the classical, single spin-wave result (\ref{eq:geophase1spinwave}).
Thus, the general solution of the Bogoliubov-de Gennes equations (\ref{eq:BdG_ring}) is in the adiabatic limit
and for $\xi\neq 0,\frac{L}{2}$ given by
\begin{eqnarray}
g^R(\xi,0;\omega) 
&=&
\sum_{p=\pm}
e^{ -\left[ 1 - \delta\varphi_p(\omega) \right] \psi(\xi) } 
\nonumber\\
&&
\times
\left[
A_{p,1}(\omega) e^{ i k_p(\omega) \xi } +
A_{p,2}(\omega) e^{ -i k_p(\omega) \xi }
\right] ,
\nonumber\\\\
\tilde{g}^R(\xi,0;\omega) 
&=&
\sum_{p=\pm} v_p(\omega)
e^{ \left[ 1 + \delta\varphi_p(\omega) \right] \psi(\xi) } 
\nonumber\\
&&
\times
\left[
A_{p,1}(\omega) e^{ i k_p(\omega) \xi } +
A_{p,2}(\omega) e^{ -i k_p(\omega) \xi }
\right] ,
\nonumber\\
\end{eqnarray}
where
$ v_\pm = - \frac{ K_1 - K_2 }{2 M_s} 
\left[
\frac{\omega}{\gamma} \pm \sqrt{ \frac{ \left( K_1 - K_2 \right)^2 }{ 4 M_s^2 } + \frac{\omega^2}{\gamma^2} }  
\right]^{-1} $
to zeroth order in $\psi'$.
The remaining coefficients $A_{\pm,1/2}(\omega)$ are determined by the boundary conditions in the regions $0<\xi<\frac{L}{2}$ and $\frac{L}{2}<\xi<L$ that are detailed in the Appendix. The final result for the Green's functions is
\begin{widetext}
\begin{subequations} \label{eq:g_sol}
\begin{eqnarray}
g^R\left( \frac{L}{2}, 0; \omega \right) 
&=& 
\left[ 1 + 
\frac{ \frac{\omega}{\gamma} }{ \sqrt{ \frac{ \left( K_1 - K_2 \right)^2 }{ 4 M_s^2 } + \frac{\omega^2}{\gamma^2} }  } 
\right] 
\frac{ 
2 \frac{J_s}{M_s} k_+(\omega) \cos\left[ \pi \delta\varphi_+(\omega) \right] \sin\left[ \frac{ k_+(\omega) L }{2} \right] }
{ D\left(k_+(\omega),\omega\right) }
\nonumber\\
&&
+
\left[ 1 - 
\frac{ \frac{\omega}{\gamma} }{ \sqrt{ \frac{ \left( K_1 - K_2 \right)^2 }{ 4 M_s^2 } + \frac{\omega^2}{\gamma^2} }  } 
\right] 
\frac{ 
2 \frac{J_s}{M_s} k_-(\omega) \cos\left[ \pi \delta\varphi_+(\omega) \right] \sin\left[ \frac{ k_-(\omega) L }{2} \right] }
{ D\left(k_-(\omega),\omega\right) } ,
\\
\tilde{g}^R\left( \frac{L}{2}, 0; \omega \right) 
&=& 
- \frac{ K_1 - K_2 }{ 2 M_s \sqrt{ \frac{ \left( K_1 - K_2 \right)^2 }{ 4 M_s^2 } + \frac{\omega^2}{\gamma^2} }  }  
\frac{ 
2 \frac{J_s}{M_s} k_+(\omega) \cos\left[ \pi \delta\varphi_+(\omega) \right] \sin\left[ \frac{ k_+(\omega) L }{2} \right] }
{ D\left(k_+(\omega),\omega\right) }
\nonumber\\
&&
+
\frac{ K_1 - K_2 }{ 2 M_s \sqrt{ \frac{ \left( K_1 - K_2 \right)^2 }{ 4 M_s^2 } + \frac{\omega^2}{\gamma^2} }  } 
\frac{ 
\frac{J_s}{M_s} k_-(\omega) \cos\left[ \pi \delta\varphi_+(\omega) \right] \sin\left[ \frac{ k_-(\omega) L }{2} \right] }
{ D\left(k_-(\omega),\omega\right) } ,
\end{eqnarray}
where
\begin{equation}
D(k,\omega) =
\left[ 4 \left( \frac{J_s}{M_s} k \right)^2 + \left( \alpha^{\rm sp} \frac{\omega}{\gamma} \right)^2 \right] \cos\left( k L \right)
-4 \left( \frac{J_s}{M_s} k \right)^2 \cos\left[ 2\pi \delta\varphi_+(\omega) \right]
- \alpha^{\rm sp} \frac{\omega}{\gamma} \left[
\alpha^{\rm sp} \frac{\omega}{\gamma} + 4 i \frac{J_s}{M_s} k \sin\left( k L \right) \right]  . 
\end{equation}
\end{subequations}
\end{widetext}
From the above solution (\ref{eq:g_sol}) for the Green's functions,
we find that the effect of the geometric phase on magnon transport through the ring is twofold:
First, as long as the interface damping enhancement 
$ \alpha^{\rm sp} \frac{\omega}{\gamma} $ is small compared to 
$ \frac{J_s}{M_s} k $, which is the typical experimental situation,
the magnon scattering resonances will be split from $kL = 2\pi n$
to $ kL = 2\pi \left[ n \pm \delta\varphi_+(\omega) \right] $,
where $n$ is an integer.
Because there is no similar phase shift in the 
$\sin\left[ \frac{ k_+(\omega) L }{2} \right]$ 
term in the numerator,
transmission at the original resonances $kL=2\pi n$ is strongly suppressed.
Second, the prefactor of 
$ \cos\left[ \pi \delta\varphi_+(\omega) \right] $
leads to additional destructive interference when $\delta\varphi_+(\omega) < 1$,
i.e., for frequencies $|\omega|$ on the order of $ \gamma \left| K_1 - K_2 \right| / 2 M_s $.
Both effects are clearly visible in Fig.~\ref{fig:transmission} which shows the result for the transmission function as a function of frequency. The transmission function at a fixed frequency determines the propagation of a single spin wave at that frequency. This implies that for single coherent spin waves, the geometric phases that occur via the position dependent anisotropies in the device in Fig.~\ref{fig:ring} can lead to --- depending on the energy --- destructive interference between spin waves that propagate clockwise and counterclockwise between the two leads. For a homogeneous anisotropy this destructive interference would be absent. If one would be able to experimentally switch between the inhomogeneous and homogeneous anistropy, one could switch the propagation of spin waves at some specific energies on or off. 

\begin{figure} 
\includegraphics[width=1\columnwidth]{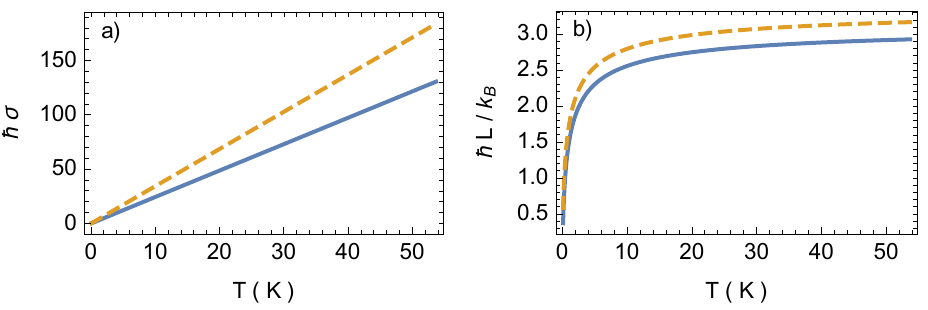} 
\caption{
a) Spin conductance (\ref{eq:sigma}) and 
b) sin Seebeck conductance (\ref{eq:kappa})
as functions of temperature,
with (solid lines) and without (dashed lines) geometric phase.
Parameters are the same as in Fig.~\ref{fig:transmission}.
}
\label{fig:conductances}
\end{figure}

For incoherent magnons, the effects resulting from the geometric phase decrease the overall magnon conductance as we discuss now. When the differences in temperature between the two leads 
and their respective spin accumulations are small
compared to the average temperature $T = \frac{1}{2} \left( T_1 + T_2 \right) $,
we may approximate the spin current (\ref{eq:current}) as
\begin{equation}
I_{1\to 2} = 
\sigma \left( \mu_2 - \mu_1 \right) +
L \left( T_2 - T_1 \right) ,
\end{equation}
with the spin and spin Seebeck conductances
\begin{subequations}
\begin{align}
\sigma =
&
\frac{1}{ k_B T } \int_{-\infty}^\infty \frac{ d \omega }{ 2\pi } 
\frac{ {\cal T}_{ 1 \to 2 }(\omega) }{ 4 \sinh^2\left( \frac{ \hbar \omega }{ 2 k_B T } \right) } ,
\label{eq:sigma}
\\  
L =
&
\frac{\hbar}{ k_B T^2 } \int_{-\infty}^\infty \frac{ d \omega }{ 2\pi } 
\frac{ \omega {\cal T}_{ 1 \to 2 }(\omega) }{ 4 \sinh^2\left( \frac{ \hbar \omega }{ 2 k_B T } \right) } .
\label{eq:kappa}
\end{align}
\end{subequations}
Plots of these conductances with and without geometric phase effects are shown in Fig.~\ref{fig:conductances}
as function of temperature.
While the geometric phase does not add any new qualitative features to these conductances,
it leads to a overall decrease,
in agreement with the preceding discussion. Because the magnon conductances are the result of averaging over a thermal magnon distribution, the geometric-phase effects are not as prominent as for a single spin wave. The fact that the energy scales set by anisotropies are typically much smaller than the thermal energy implies that the geometric phases resulting from anistropy will typically not strongly affect the magnon transport.

\section{Conclusion, discussion, and outlook}
In this article we have discussed the geometric phases that arise in semi-classical magnetic dynamics using the framework developed by Hannay \cite{Hannay_1985}. In doing so, we have, as part of our dicussion, rederived some known results for the so-called magnon Berry phase \cite{PhysRevB.72.024456}, as well as the geometric phase of quasi-particles in a superfluid \cite{PhysRevLett.97.040401}. Finally, we have developed a framework for magnon transport in heterostructures consisting of metallic leads connected to a ferromagnetic insulator, that incorporates magnon ellipticity. This formalism is therefore able to incorporate the Hannay angle due to varying magnon or spin-wave ellipticity, and we have applied it to a simple device. We found that, while the Hannay angle due to position-dependent ellipticity could be used to engineer destructive interference between two coherent spin waves, it does not dramatically affect the transport of thermal magnons. 

Experimentally engineering time- or position-dependent anisotropies may be challenging. The latter can perhaps be achieved by varying the thickness of a film of material with a perpendicular magnetic interface anisotropy \cite{RevModPhys.89.025006}, such that for small thicknesses the anisotropy is out-of-plane, whereas for larger thickness it becomes in-plane due to magnetostatic effects. Time-dependent anisotropies could perhaps be achieved via magnetoelastic effects and the application of time-dependent strain \cite{PhysRevLett.120.257203}.

Let us mention several possible extensions and follow-ups of the work presented here: First, one could consider a general micromagnetic energy functional, rather than the specific toy model that we focused on for pedagogical purposes. One would then proceed by finding, at each instance in time, the equilibrium magnetization direction $\vm_0 (t)$. Subsequently, one would have to apply the transformation of Sec.~\ref{subsubsec:ellipsetdptfield} with $\vn$ replaced by $\vm_0 (t)$ to proceed along the same lines as highlighted in that section. While straightforward, these developments can be somewhat cumbersome because $\vm_0 (t)$, and thus the angles $\phi$ and $\theta$, as well as the energy that describes quadratic fluctuations around $\vm_0 (t)$, all have to be found from minimizing the energy and expanding around this minimum at each time $t$. Moreover, an arbitrary adiabatic variation of the parameters in the energy does not need to conserve the energy so that the area of the ellipse of precession changes. The action variable $r(t)$ then needs to be chosen differently to accomodate for this change in area. A second interesting generalization of the theory presented here would be to consider spatially two-dimensional situations in which one would expect Hall-like effects in, e.g., the magnon transport \cite{PhysRevLett.106.197202}. Finally, the discussion of the Hannay angles that occur in single-domain magnetization dynamics and for spin waves could be extended to the nonlinear regime provided damping is neglected. In view of this, a useful development would be to extend the theory to the case of spin-torque nano-oscillators \cite{KIM2012217} as well. This would, however, require the proper inclusion of dissipative effects, which is beyond the scope of this paper. 

Finally, we mention that most of the approaches presented here can be extended to other systems involving semi-classical spin dynamics. These include antiferromagnets \cite{10754/627305}, magnetoelastic systems \cite{Foerster_2019}, and magnetomechanical systems \cite{losby2016spin}. We hope that this article provides a useful starting point for undertaking such excursions.

\acknowledgments
It is a pleasure to thank Miguel Lammers for doing his bachelor research project on this topic, and Alexander Serga for discussions. This work is supported by the European Research Council via Consolidator
Grant number 725509 SPINBEYOND. RD is member of the D-ITP consortium, a program of the
Netherlands Organisation for Scientific Research (NWO) that is funded by the Dutch Ministry of Education, Culture and Science (OCW).

\section*{Appendix: Non-equilibrium Green's functions for elliptical magnons}
\label{app:Green}

\subsection{Derivation of Bogoliubov-de Gennes equations}

In this Appendix, we briefly outline the main ingredients
of the non-equilibrium Green's function technique for elliptical magnons
that we used to obtain the spin current (\ref{eq:current}).
For that purpose, it is convenient to write the magnon Hamiltonian (\ref{eq:H_boson})
in the form
\begin{equation}
{\cal H}_\psi^{\rm ex} = \frac{1}{2} \int d\xi \int d\xi' 
\left( a^\dagger(\xi) , a(\xi) \right)
\hat{\cal H}(\xi,\xi') 
\left( \begin{matrix}
a(\xi') \\ a^\dagger(\xi')
\end{matrix} \right) ,
\end{equation}
with a $2\times 2$ Hamiltonian matrix $\hat{\cal H}(\xi,\xi')$.
Similarly, we collect the normal and anomalous retarded magnon Green's functions
into a $2\times 2$ matrix:
\begin{equation}
\hat{G}^R(\xi,t;\xi',t') =
\left( \begin{matrix}
g^R(\xi,t;\xi',t') & \left[ \tilde{g}^R(\xi,t;\xi',t') \right]^* \\
\tilde{g}^R(\xi,t;\xi',t') & \left[ g^R(\xi,t;\xi',t') \right]^*
\end{matrix} \right) .
\end{equation}
The microscopic definition of these retarded Green's functions is
\begin{subequations} \label{eq:gR}
	\begin{eqnarray}
	g^R(\xi,t;\xi',t') 
	&=& -\frac{i}{\hbar}\Theta(t-t')
	\left\langle \left[ a(\xi,t), a^\dagger(\xi',t') \right] \right\rangle , 
	\nonumber\\\\
	\tilde{g}^R(\xi,t;\xi',t') 
	&=& -\frac{i}{\hbar}\Theta(t-t')
	\left\langle \left[ a^\dagger(\xi,t), a^\dagger(\xi',t') \right] \right\rangle . 
	\nonumber\\
	\end{eqnarray}
\end{subequations}
The corresponding advanced Green's function matrix is
$ \hat{G}^A(\xi,t;\xi',t') = \left[ \hat{G}^R(\xi',t';\xi,t) \right]^\dagger $.
Retarded and advanced Green's function matrices satisfy the
Dyson equation \cite{rammer2004quantum}
\begin{align} \label{eq:Dyson_RA}
&\hat{\sigma}_3 i\hbar \partial_t \hat{G}^{R/A}(\xi,t;\xi',t') -
\int d\xi_1 \hat{\cal H}(\xi,\xi_1) \hat{G}^{R/A}(\xi_1,t;\xi',t') 
\nonumber\\
&
=
\int dt_1 \int d\xi_1 \hat{\Sigma}^{R/A}(\xi,t;\xi_1,t_1) \hat{G}^{R/A}(\xi_1,t_1;\xi',t') ,  
\end{align}
with the retarded and advanced self-energies $\hat{\Sigma}^{R/A}(\xi,t;\xi',t')$,
and the Pauli matrix $\hat{\sigma}_3 = {\rm diag}(1,-1)$.
While retarded and advanced Green functions contain the information about the magnon propagation,
the magnon distribution is encoded in the normal and anomalous Keldysh Green's functions: 
\begin{subequations}
\begin{align}
g^K(\xi,t;\xi',t') &=  - \frac{i}{\hbar} \left\langle \left\{
a(\xi,t) , a^\dagger (\xi',t')
\right\} \right\rangle ,
\\
\tilde{g}^K(\xi,t;\xi',t') &=  - \frac{i}{\hbar} \left\langle \left\{
a^\dagger(\xi,t) , a^\dagger (\xi',t')
\right\} \right\rangle ,
\end{align}
\end{subequations}
where $\{\cdot,\cdot\}$ is the anticommutator.
Collecting the Keldysh Green's functions into a matrix as well,
\begin{equation}
\hat{G}^K(\xi,t;\xi',t') =
\left( \begin{matrix}
g^K(\xi,t;\xi',t') & -\left[ \tilde{g}^K(\xi,t;\xi',t') \right]^* \\
\tilde{g}^K(\xi,t;\xi',t') & -\left[ g^K(\xi,t;\xi',t') \right]^*
\end{matrix} \right) ,
\end{equation}
one finds that they obey the following Dyson equations \cite{rammer2004quantum}:
\begin{subequations} \label{eq:Dyson_K}
\begin{align}
&\hat{\sigma}_3 i\hbar \partial_t \hat{G}^{K}(\xi,t;\xi',t') -
\int d\xi_1 \hat{\cal H}(\xi,\xi_1) \hat{G}^{K}(\xi_1,t;\xi',t') 
\nonumber\\
=
&
\int dt_1 \int d\xi_1 \biggl[ 
\hat{\Sigma}^{K}(\xi,t;\xi_1,t_1) \hat{G}^{A}(\xi_1,t_1;\xi',t') 
\nonumber\\
& \phantom{\int dt_1 \int d\xi_1 \biggl[ }
+
\hat{\Sigma}^{R}(\xi,t;\xi_1,t_1) \hat{G}^{K}(\xi_1,t_1;\xi',t') 
\biggr] ,  
\\
&-\hat{\sigma}_3 i\hbar \partial_{t'} \hat{G}^{K}(\xi,t;\xi',t') -
\int d\xi_1 \hat{G}^{K}(\xi,t;\xi_1,t') \hat{\cal H}(\xi_1,\xi')  
\nonumber\\
=
&
\int dt_1 \int d\xi_1 \biggl[ 
\hat{G}^{R}(\xi,t;\xi_1,t_1) \hat{\Sigma}^{K}(\xi_1,t_1;\xi',t') 
\nonumber\\
& \phantom{\int dt_1 \int d\xi_1 \biggl[ }
+
\hat{G}^{K}(\xi,t;\xi_1,t_1) \hat{\Sigma}^{A}(\xi_1,t_1;\xi',t') 
\biggr] ,  
\end{align}
\end{subequations}
where $\hat{\Sigma}^K(\xi,t;\xi',t')$ is the Keldysh self-energy.
Provided the self-energies are known in some approximation,
the Dyson equations (\ref{eq:Dyson_RA}) and (\ref{eq:Dyson_K})
contain all information about the spectrum and distribution
of single-magnon excitations.

The local spin density can be expressed via the Keldysh Green's function as 
\begin{align}
s(\xi,t) 
&= s - \left\langle a^\dagger(\xi,t) a(\xi,t) \right\rangle 
\\
&= s + \frac{1}{2} \delta(0) - \frac{i\hbar}{4} {\rm Tr}\, \hat{G}^K(\xi,t;\xi,t) ,
\end{align}
where the $\delta(0)$ term should be regularized as $a_0^{-1}$,
with the microscopic lattice constant $a_0$.
From the Keldysh Dyson equations (\ref{eq:Dyson_K}),
we thus obtain the equation of motion of the spin density:
\begin{widetext}
\begin{align}
\partial_t s(\xi,t) =
& 
\frac{1}{4} {\rm Tr} \int d\xi_1 \left[
\hat{\sigma}_3 \hat{\cal H}(\xi,\xi_1) \hat{G}^K(\xi_1,t;\xi,t) -
\hat{G}^K(\xi,t;\xi_1,t) \hat{\cal H}(\xi_1,\xi) \hat{\sigma}_3
\right]
\nonumber\\
&
+ \frac{1}{4} {\rm Tr} \int dt_1 \int d\xi_1 \left[
\hat{\sigma}_3 \hat{\Sigma}^K(\xi,t;\xi_1,t_1) \hat{G}^A(\xi_1,t_1;\xi,t) -
\hat{G}^R(\xi,t;\xi_1,t_1) \hat{\Sigma}^K(\xi_1,t_1;\xi,t) \hat{\sigma}_3 
\right]
\nonumber\\
&
+ \frac{1}{4} {\rm Tr} \int dt_1 \int d\xi_1 \left[
\hat{\sigma}_3 \hat{\Sigma}^R(\xi,t;\xi_1,t_1) \hat{G}^K(\xi_1,t_1;\xi,t) -
\hat{G}^K(\xi,t;\xi_1,t_1) \hat{\Sigma}^A(\xi_1,t_1;\xi,t) \hat{\sigma}_3 
\right] .
\end{align}
For a magnet in contact with two leads,
we may split the self-energies into bulk and lead contributions,
\begin{equation}
\hat{\Sigma}^{R/A/K}(\xi,t;\xi',t') =
\hat{\Sigma}^{R/A/K}_{\rm bulk}(\xi,t;\xi',t') +
\hat{\Sigma}^{R/A/K}_1(\xi,t;\xi',t') +
\hat{\Sigma}^{R/A/K}_2(\xi,t;\xi',t') ,
\end{equation}
where $1$ and $2$ refers to the two leads.
The total spin lost or gained at lead $2$ is consequently given by
\begin{align}
I_2 (t)
=
& 
\frac{1}{4} {\rm Tr} \int dt_1 \int d\xi \int d\xi_1\left[
\hat{\sigma}_3 \hat{\Sigma}^K_2(\xi,t;\xi_1,t_1) \hat{G}^A(\xi_1,t_1;\xi,t) -
\hat{G}^R(\xi,t;\xi_1,t_1) \hat{\Sigma}^K_2(\xi_1,t_1;\xi,t) \hat{\sigma}_3 
\right]
\nonumber\\
&
+ \frac{1}{4} {\rm Tr} \int dt_1 \int d\xi \int d\xi_1 \left[
\hat{\sigma}_3 \hat{\Sigma}^R_2(\xi,t;\xi_1,t_1) \hat{G}^K(\xi_1,t_1;\xi,t) -
\hat{G}^K(\xi,t;\xi_1,t_1) \hat{\Sigma}^A_2(\xi_1,t_1;\xi,t) \hat{\sigma}_3 
\right] .
\label{eq:I_2}
\end{align}
In a steady state, this reduces to
\begin{align}
I_2
=
& 
\frac{1}{4} {\rm Tr} \int \frac{d\omega}{2\pi} \int d\xi \int d\xi_1 \left[
\hat{\sigma}_3 \hat{\Sigma}^K_2(\xi,\xi_1;\omega) \hat{G}^A(\xi_1,\xi;\omega) -
\hat{G}^R(\xi,\xi_1;\omega) \hat{\Sigma}^K_2(\xi_1,\xi;\omega) \hat{\sigma}_3 
\right]
\nonumber\\
&
+ \frac{1}{4} {\rm Tr} \int \frac{d\omega}{2\pi} \int d\xi \int d\xi_1 \left[
\hat{\sigma}_3 \hat{\Sigma}^R_2(\xi,\xi_1;\omega) \hat{G}^K(\xi_1,\xi;\omega) -
\hat{G}^K(\xi,\xi_1;\omega) \hat{\Sigma}^A_2(\xi_1,\xi;\omega) \hat{\sigma}_3 
\right] .
\label{eq:I_2_omega}
\end{align}
In this case,
we can furthermore directly solve the Keldysh Dyson equations (\ref{eq:Dyson_K})
in frequency space, yielding
\begin{align}
\hat{G}^K(\xi,\xi';\omega) =
&
\int d\xi_1 \int d\xi_2 \hat{G}^R(\xi,\xi_1;\omega) \hat{\Sigma}^K(\xi_1,\xi_2;\omega) \hat{G}^A(\xi_2,\xi';\omega) .
\end{align}
The retarded and advanced self-energies corresponding to bulk Gilbert damping and spin pumping from the electronic leads
can be obtained from the Landau-Lifshitz-Gilbert phenomenology \cite{PhysRevB.96.174422};
they are
\begin{subequations}
\begin{align}
\hat{\Sigma}^{R/A}_{\rm bulk}(\xi,\xi';\omega) 
&
= \mp i \alpha \hbar \omega
\left(\begin{matrix}
1 & 0 \\
0 & 1 
\end{matrix} \right)
\delta(\xi-\xi')  ,
\\
\hat{\Sigma}^{R/A}_1(\xi,\xi';\omega)
&
= \mp i \alpha_1^{\rm sp}
\left( \begin{matrix}
\hbar \omega - \mu_1 & 0 \\
0 & \hbar \omega + \mu_1
\end{matrix} \right)
\delta(\xi) \delta(\xi-\xi') ,
\\
\hat{\Sigma}^{R/A}_2(\xi,\xi';\omega)
&
= \mp i \alpha_2^{\rm sp}
\left( \begin{matrix}
\hbar \omega - \mu_2 & 0 \\
0 & \hbar \omega + \mu_2
\end{matrix} \right)
\delta\left(\xi - \frac{L}{2} \right) \delta(\xi-\xi') ,
\end{align}
\end{subequations}
The Keldysh self-energies are then determined by the fluctuation-dissipation theorem,
so that
\begin{subequations}
\begin{align}
\hat{\Sigma}^{K}_{\rm bulk}(\xi,\xi';\omega) 
&
= -2 i \alpha \hbar \omega \coth\left( \frac{\hbar\omega}{2 k_B T} \right) 
\left(\begin{matrix}
1 & 0 \\
0 & 1 
\end{matrix} \right)
\delta(\xi-\xi') ,
\\
\hat{\Sigma}^{K}_1(\xi,\xi';\omega)
&
= -2 i \alpha_1^{\rm sp}
\left( \begin{matrix}
\left( \hbar \omega - \mu_1 \right) \coth\left( \frac{ \hbar \omega - \mu_1 }{2 k_B T_1} \right) & 0 \\
0 & \left( \hbar \omega + \mu_1 \right) \coth\left( \frac{ \hbar \omega + \mu_1 }{2 k_B T_1} \right)
\end{matrix} \right)
\delta(\xi) \delta(\xi-\xi') ,
\\
\hat{\Sigma}^{K}_2(\xi,\xi';\omega)
&
= -2 i \alpha_2^{\rm sp}
\left( \begin{matrix}
\left( \hbar \omega - \mu_2 \right) \coth\left( \frac{ \hbar \omega - \mu_2 }{2 k_B T_2} \right) & 0 \\
0 & \left( \hbar \omega + \mu_2 \right) \coth\left( \frac{ \hbar \omega + \mu_2 }{2 k_B T_2} \right)
\end{matrix} \right)
\delta\left(\xi - \frac{L}{2} \right) \delta(\xi-\xi') ,
\end{align}
\end{subequations}
With these self-energies the Dyson equation (\ref{eq:Dyson_RA}) 
reduces to the Bogoliubov-de Gennes equations 
	\begin{subequations} \label{eq:BdG_ring}
		\begin{eqnarray}
		&&
		\left[ 
		\frac{1+i\alpha}{\gamma} \omega  +\frac{J_s}{M_s} \partial_\xi^2 - \mu_0 H_{\rm ext} - \frac{ K_1 + K_2 }{ 2 M_s } 
		+ \delta(\xi) \frac{ i \alpha_1^{\rm sp} }{ \hbar\gamma } \left( \hbar\omega - \mu_1 \right)
		+ \delta\left(\xi-\frac{L}{2} \right) \frac{ i \alpha_2^{\rm sp} }{ \hbar\gamma } \left( \hbar\omega - \mu_2 \right)
		\right] 
		g^R(\xi,0;\omega)
		\nonumber\\
		&&
		-\frac{ K_1 - K_2 }{ 2 M_s } e^{-2 i \psi(\xi)} \tilde{g}^R(\xi,0;\omega)
		= \frac{1}{\hbar\gamma}\delta(\xi) , 
		\nonumber\\\\
		&&
		\left[ 
		-\frac{1-i\alpha}{\gamma} \omega  +\frac{J_s}{M_s} \partial_\xi^2 - \mu_0 H_{\rm ext} - \frac{ K_1 + K_2 }{ 2 M_s } 
		+ \delta(\xi) \frac{ i \alpha_1^{\rm sp} }{ \hbar\gamma } \left( \hbar\omega + \mu_1 \right)
		+ \delta\left(\xi-\frac{L}{2} \right) \frac{ i \alpha_2^{\rm sp} }{ \hbar\gamma } \left( \hbar\omega + \mu_2 \right)
		\right] 
		\tilde{g}^R(\xi,0;\omega)
		\nonumber\\
		&&
		-\frac{ K_1 - K_2 }{ 2 M_s } e^{2 i \psi(\xi)} g^R(\xi,0;\omega)
		= 0 . 
		\nonumber\\
		\end{eqnarray}
	\end{subequations}
\end{widetext}
These are solved in Sec.~\ref{sec:magnons}. We derive the spin current (\ref{eq:current}) from lead $1$ to lead $2$   
by isolating the contribution of lead $1$ 
to the total spin lost or gained at lead $2$, see Eq.~(\ref{eq:I_2_omega}). 

\subsection{Boundary conditions}

The solutions on the ring have to be periodic, hence
$ g^R(0,0;\omega) = g^R(L,0;\omega) $ and
$ \tilde{g}^R(0,0;\omega) = \tilde{g}^R(L,0;\omega) $.
Furthermore, they must be continuous at $\xi=\frac{L}{2}$, i.e.,
$ g^R\left(\frac{L}{2}+0^+,0;\omega\right) = g^R\left(\frac{L}{2}-0^+,0;\omega\right) $ and
$ \tilde{g}^R(\frac{L}{2}+0^+,0;\omega) = \tilde{g}^R(\frac{L}{2}-0^+,0;\omega) $.
Integrating the Bogoliubov-de Gennes equations (\ref{eq:BdG_ring}) 
from $\xi=\frac{L}{2}-0^+$ to $\xi=\frac{L}{2}+0^+$ and
from $\xi=-0^+=L$ to $\xi=0^+$ additionally yields the following spin pumping boundary conditions:
\begin{subequations}
	\begin{eqnarray}
	&&
	\frac{J_s}{M_s} \left[ 
	\partial_\xi g^R\left(\xi,0;\omega\right) \bigl|_{\xi=0} - \partial_\xi g^R\left(\xi,0;\omega\right) \bigl|_{\xi=L}
	\right]
	\nonumber\\
	&&
	+ \frac{ i \alpha_1^{\rm sp} }{ \hbar\gamma } \left( \hbar\omega - \mu_1 \right) g^R\left(0,0;\omega\right) 
	=\frac{1}{\hbar\gamma},
	\\
	&&
	\frac{J_s}{M_s} \left[ 
	\partial_\xi g^R\left(\xi,0;\omega\right) \bigl|_{\xi=\frac{L}{2}+0^+} - 
	\partial_\xi g^R\left(\xi,0;\omega\right) \bigl|_{\xi=\frac{L}{2}-0^+}
	\right]
	\nonumber\\
	&&
	+ \frac{ i \alpha_2^{\rm sp} }{ \hbar\gamma } \left( \hbar\omega - \mu_2 \right) g^R\left(\frac{L}{2},0;\omega\right) 
	=0,
	\end{eqnarray}
\end{subequations}
and
\begin{subequations}
	\begin{eqnarray}
	&&
	\frac{J_s}{M_s} \left[ 
	\partial_\xi \tilde{g}^R\left(\xi,0;\omega\right) \bigl|_{\xi=0} - \partial_\xi \tilde{g}^R\left(\xi,0;\omega\right) \bigl|_{\xi=L}
	\right]
	\nonumber\\
	&&
	+ \frac{ i \alpha_1^{\rm sp} }{ \hbar\gamma } \left( \hbar\omega + \mu_1 \right) \tilde{g}^R\left(0,0;\omega\right) 
	=0,
	\\
	&&
	\frac{J_s}{M_s} \left[ 
	\partial_\xi \tilde{g}^R\left(\xi,0;\omega\right) \bigl|_{\xi=\frac{L}{2}+0^+} - 
	\partial_\xi \tilde{g}^R\left(\xi,0;\omega\right) \bigl|_{\xi=\frac{L}{2}-0^+}
	\right]
	\nonumber\\
	&&
	+ \frac{ i \alpha_2^{\rm sp} }{ \hbar\gamma } \left( \hbar\omega + \mu_2 \right) \tilde{g}^R\left(\frac{L}{2},0;\omega\right) 
	=0.
	\end{eqnarray}
\end{subequations}
Solving these boundary conditions in the limit $|\mu_{1/2}| \ll |\hbar\omega|$
and for $\alpha_1^{\rm sp}=\alpha_2^{\rm sp}=\alpha^{\rm sp}$,
we obtain analytical expressions for the relevant Green's functions [Eq.~(\ref{eq:g_sol})].

\bibliography{main}

\begin{thebibliography}{29}
\expandafter\ifx\csname natexlab\endcsname\relax\def\natexlab#1{#1}\fi
\expandafter\ifx\csname bibnamefont\endcsname\relax
  \def\bibnamefont#1{#1}\fi
\expandafter\ifx\csname bibfnamefont\endcsname\relax
  \def\bibfnamefont#1{#1}\fi
\expandafter\ifx\csname citenamefont\endcsname\relax
  \def\citenamefont#1{#1}\fi
\expandafter\ifx\csname url\endcsname\relax
  \def\url#1{\texttt{#1}}\fi
\expandafter\ifx\csname urlprefix\endcsname\relax\def\urlprefix{URL }\fi
\providecommand{\bibinfo}[2]{#2}
\providecommand{\eprint}[2][]{\url{#2}}

\bibitem[{\citenamefont{Hannay}(1985)}]{Hannay_1985}
\bibinfo{author}{\bibfnamefont{J.~H.} \bibnamefont{Hannay}},
  \bibinfo{journal}{Journal of Physics A: Mathematical and General}
  \textbf{\bibinfo{volume}{18}}, \bibinfo{pages}{221} (\bibinfo{year}{1985}),
  \urlprefix\url{https://doi.org/10.1088%2F0305-4470%2F18%2F2%2F011}.

\bibitem[{\citenamefont{Chruscinski and
  Jamiolkowski}(2004)}]{chruscinski2004geometric}
\bibinfo{author}{\bibfnamefont{D.}~\bibnamefont{Chruscinski}} \bibnamefont{and}
  \bibinfo{author}{\bibfnamefont{A.}~\bibnamefont{Jamiolkowski}},
  \emph{\bibinfo{title}{Geometric Phases in Classical and Quantum Mechanics}},
  Progress in Mathematical Physics (\bibinfo{publisher}{Birkh{\"a}user Boston},
  \bibinfo{year}{2004}), ISBN \bibinfo{isbn}{9780817642822},
  \urlprefix\url{https://www.springer.com/gp/book/9780817642822}.

\bibitem[{\citenamefont{Berry}(1984)}]{berry1984quantal}
\bibinfo{author}{\bibfnamefont{M.~V.} \bibnamefont{Berry}},
  \bibinfo{journal}{Proc. R. Soc. Lond. A} \textbf{\bibinfo{volume}{392}},
  \bibinfo{pages}{45} (\bibinfo{year}{1984}),
  \urlprefix\url{http://rspa.royalsocietypublishing.org/content/392/1802/45}.

\bibitem[{\citenamefont{Resta and Vanderbilt}(2007)}]{Resta2007}
\bibinfo{author}{\bibfnamefont{R.}~\bibnamefont{Resta}} \bibnamefont{and}
  \bibinfo{author}{\bibfnamefont{D.}~\bibnamefont{Vanderbilt}},
  \emph{\bibinfo{title}{Theory of Polarization: A Modern Approach}}
  (\bibinfo{publisher}{Springer Berlin Heidelberg}, \bibinfo{address}{Berlin,
  Heidelberg}, \bibinfo{year}{2007}), pp. \bibinfo{pages}{31--68}, ISBN
  \bibinfo{isbn}{978-3-540-34591-6},
  \urlprefix\url{https://doi.org/10.1007/978-3-540-34591-6_2}.

\bibitem[{\citenamefont{Xiao et~al.}(2010)\citenamefont{Xiao, Chang, and
  Niu}}]{RevModPhys.82.1959}
\bibinfo{author}{\bibfnamefont{D.}~\bibnamefont{Xiao}},
  \bibinfo{author}{\bibfnamefont{M.-C.} \bibnamefont{Chang}}, \bibnamefont{and}
  \bibinfo{author}{\bibfnamefont{Q.}~\bibnamefont{Niu}}, \bibinfo{journal}{Rev.
  Mod. Phys.} \textbf{\bibinfo{volume}{82}}, \bibinfo{pages}{1959}
  (\bibinfo{year}{2010}),
  \urlprefix\url{https://link.aps.org/doi/10.1103/RevModPhys.82.1959}.

\bibitem[{\citenamefont{Nagaosa et~al.}(2010)\citenamefont{Nagaosa, Sinova,
  Onoda, MacDonald, and Ong}}]{RevModPhys.82.1539}
\bibinfo{author}{\bibfnamefont{N.}~\bibnamefont{Nagaosa}},
  \bibinfo{author}{\bibfnamefont{J.}~\bibnamefont{Sinova}},
  \bibinfo{author}{\bibfnamefont{S.}~\bibnamefont{Onoda}},
  \bibinfo{author}{\bibfnamefont{A.~H.} \bibnamefont{MacDonald}},
  \bibnamefont{and} \bibinfo{author}{\bibfnamefont{N.~P.} \bibnamefont{Ong}},
  \bibinfo{journal}{Rev. Mod. Phys.} \textbf{\bibinfo{volume}{82}},
  \bibinfo{pages}{1539} (\bibinfo{year}{2010}),
  \urlprefix\url{https://link.aps.org/doi/10.1103/RevModPhys.82.1539}.

\bibitem[{\citenamefont{Hasan and Kane}(2010)}]{RevModPhys.82.3045}
\bibinfo{author}{\bibfnamefont{M.~Z.} \bibnamefont{Hasan}} \bibnamefont{and}
  \bibinfo{author}{\bibfnamefont{C.~L.} \bibnamefont{Kane}},
  \bibinfo{journal}{Rev. Mod. Phys.} \textbf{\bibinfo{volume}{82}},
  \bibinfo{pages}{3045} (\bibinfo{year}{2010}),
  \urlprefix\url{https://link.aps.org/doi/10.1103/RevModPhys.82.3045}.

\bibitem[{\citenamefont{Qi and Zhang}(2011)}]{RevModPhys.83.1057}
\bibinfo{author}{\bibfnamefont{X.-L.} \bibnamefont{Qi}} \bibnamefont{and}
  \bibinfo{author}{\bibfnamefont{S.-C.} \bibnamefont{Zhang}},
  \bibinfo{journal}{Rev. Mod. Phys.} \textbf{\bibinfo{volume}{83}},
  \bibinfo{pages}{1057} (\bibinfo{year}{2011}),
  \urlprefix\url{https://link.aps.org/doi/10.1103/RevModPhys.83.1057}.

\bibitem[{\citenamefont{Sinitsyn}(2009)}]{Sinitsyn_2009}
\bibinfo{author}{\bibfnamefont{N.~A.} \bibnamefont{Sinitsyn}},
  \bibinfo{journal}{Journal of Physics A: Mathematical and Theoretical}
  \textbf{\bibinfo{volume}{42}}, \bibinfo{pages}{193001}
  (\bibinfo{year}{2009}),
  \urlprefix\url{https://doi.org/10.1088%2F1751-8113%2F42%2F19%2F193001}.

\bibitem[{\citenamefont{Dugaev et~al.}(2005)\citenamefont{Dugaev, Bruno,
  Canals, and Lacroix}}]{PhysRevB.72.024456}
\bibinfo{author}{\bibfnamefont{V.~K.} \bibnamefont{Dugaev}},
  \bibinfo{author}{\bibfnamefont{P.}~\bibnamefont{Bruno}},
  \bibinfo{author}{\bibfnamefont{B.}~\bibnamefont{Canals}}, \bibnamefont{and}
  \bibinfo{author}{\bibfnamefont{C.}~\bibnamefont{Lacroix}},
  \bibinfo{journal}{Phys. Rev. B} \textbf{\bibinfo{volume}{72}},
  \bibinfo{pages}{024456} (\bibinfo{year}{2005}),
  \urlprefix\url{https://link.aps.org/doi/10.1103/PhysRevB.72.024456}.

\bibitem[{\citenamefont{Matsumoto and Murakami}(2011)}]{PhysRevLett.106.197202}
\bibinfo{author}{\bibfnamefont{R.}~\bibnamefont{Matsumoto}} \bibnamefont{and}
  \bibinfo{author}{\bibfnamefont{S.}~\bibnamefont{Murakami}},
  \bibinfo{journal}{Phys. Rev. Lett.} \textbf{\bibinfo{volume}{106}},
  \bibinfo{pages}{197202} (\bibinfo{year}{2011}),
  \urlprefix\url{https://link.aps.org/doi/10.1103/PhysRevLett.106.197202}.

\bibitem[{\citenamefont{Zhang et~al.}(2006)\citenamefont{Zhang, Dudarev, and
  Niu}}]{PhysRevLett.97.040401}
\bibinfo{author}{\bibfnamefont{C.}~\bibnamefont{Zhang}},
  \bibinfo{author}{\bibfnamefont{A.~M.} \bibnamefont{Dudarev}},
  \bibnamefont{and} \bibinfo{author}{\bibfnamefont{Q.}~\bibnamefont{Niu}},
  \bibinfo{journal}{Phys. Rev. Lett.} \textbf{\bibinfo{volume}{97}},
  \bibinfo{pages}{040401} (\bibinfo{year}{2006}),
  \urlprefix\url{https://link.aps.org/doi/10.1103/PhysRevLett.97.040401}.

\bibitem[{\citenamefont{Landau and Lifschitz}(1992)}]{LANDAU199251}
\bibinfo{author}{\bibfnamefont{L.}~\bibnamefont{Landau}} \bibnamefont{and}
  \bibinfo{author}{\bibfnamefont{E.}~\bibnamefont{Lifschitz}}, in
  \emph{\bibinfo{booktitle}{Perspectives in Theoretical Physics}}, edited by
  \bibinfo{editor}{\bibfnamefont{L.}~\bibnamefont{Pitaevski}}
  (\bibinfo{publisher}{Pergamon}, \bibinfo{address}{Amsterdam},
  \bibinfo{year}{1992}), pp. \bibinfo{pages}{51 -- 65}, ISBN
  \bibinfo{isbn}{978-0-08-036364-6},
  \urlprefix\url{http://www.sciencedirect.com/science/article/pii/B9780080363646500089}.

\bibitem[{\citenamefont{Gilbert}(2004)}]{1353448}
\bibinfo{author}{\bibfnamefont{T.}~\bibnamefont{Gilbert}},
  \bibinfo{journal}{Magnetics, IEEE Transactions on}
  \textbf{\bibinfo{volume}{40}}, \bibinfo{pages}{3443} (\bibinfo{year}{2004}),
  ISSN \bibinfo{issn}{0018-9464}.

\bibitem[{\citenamefont{Kruglyak et~al.}(2010)\citenamefont{Kruglyak,
  Demokritov, and Grundler}}]{Kruglyak_2010}
\bibinfo{author}{\bibfnamefont{V.~V.} \bibnamefont{Kruglyak}},
  \bibinfo{author}{\bibfnamefont{S.~O.} \bibnamefont{Demokritov}},
  \bibnamefont{and} \bibinfo{author}{\bibfnamefont{D.}~\bibnamefont{Grundler}},
  \bibinfo{journal}{Journal of Physics D: Applied Physics}
  \textbf{\bibinfo{volume}{43}}, \bibinfo{pages}{260301}
  (\bibinfo{year}{2010}),
  \urlprefix\url{https://doi.org/10.1088%2F0022-3727%2F43%2F26%2F260301}.

\bibitem[{\citenamefont{Bruno}(2004)}]{PhysRevLett.93.247202}
\bibinfo{author}{\bibfnamefont{P.}~\bibnamefont{Bruno}},
  \bibinfo{journal}{Phys. Rev. Lett.} \textbf{\bibinfo{volume}{93}},
  \bibinfo{pages}{247202} (\bibinfo{year}{2004}),
  \urlprefix\url{https://link.aps.org/doi/10.1103/PhysRevLett.93.247202}.

\bibitem[{\citenamefont{Pitaevskii and
  Stringari}(2003)}]{pítajevskíj2003bose}
\bibinfo{author}{\bibfnamefont{L.}~\bibnamefont{Pitaevskii}} \bibnamefont{and}
  \bibinfo{author}{\bibfnamefont{S.}~\bibnamefont{Stringari}},
  \emph{\bibinfo{title}{Bose-Einstein Condensation and Superfluidity}},
  Comparative Pathobiology - Studies in the Postmodern Theory of Education
  (\bibinfo{publisher}{Clarendon Press}, \bibinfo{year}{2003}), ISBN
  \bibinfo{isbn}{9780198507192},
  \urlprefix\url{https://books.google.nl/books?id=rIobbOxC4j4C}.

\bibitem[{\citenamefont{Holstein and Primakoff}(1940)}]{PhysRev.58.1098}
\bibinfo{author}{\bibfnamefont{T.}~\bibnamefont{Holstein}} \bibnamefont{and}
  \bibinfo{author}{\bibfnamefont{H.}~\bibnamefont{Primakoff}},
  \bibinfo{journal}{Phys. Rev.} \textbf{\bibinfo{volume}{58}},
  \bibinfo{pages}{1098} (\bibinfo{year}{1940}),
  \urlprefix\url{https://link.aps.org/doi/10.1103/PhysRev.58.1098}.

\bibitem[{\citenamefont{Cornelissen et~al.}(2015)\citenamefont{Cornelissen,
  Liu, Duine, {Ben Youssef}, and {van
  Wees}}}]{ab9817b8a4964a7ebb40e9945e0961de}
\bibinfo{author}{\bibfnamefont{L.}~\bibnamefont{Cornelissen}},
  \bibinfo{author}{\bibfnamefont{J.}~\bibnamefont{Liu}},
  \bibinfo{author}{\bibfnamefont{R.}~\bibnamefont{Duine}},
  \bibinfo{author}{\bibfnamefont{J.}~\bibnamefont{{Ben Youssef}}},
  \bibnamefont{and} \bibinfo{author}{\bibfnamefont{B.}~\bibnamefont{{van
  Wees}}}, \bibinfo{journal}{Nature Physics} \textbf{\bibinfo{volume}{11}},
  \bibinfo{pages}{1022} (\bibinfo{year}{2015}), ISSN \bibinfo{issn}{1745-2481}.

\bibitem[{\citenamefont{Rammer}(2004)}]{rammer2004quantum}
\bibinfo{author}{\bibfnamefont{J.}~\bibnamefont{Rammer}},
  \emph{\bibinfo{title}{Quantum Transport Theory}}, Frontiers in Physics
  (\bibinfo{publisher}{Avalon Publishing}, \bibinfo{year}{2004}), ISBN
  \bibinfo{isbn}{9780813346229},
  \urlprefix\url{https://www.crcpress.com/Quantum-Transport-Theory/Rammer/p/book/9780813342849}.

\bibitem[{\citenamefont{Zheng et~al.}(2017)\citenamefont{Zheng, Bender,
  Armaitis, Troncoso, and Duine}}]{PhysRevB.96.174422}
\bibinfo{author}{\bibfnamefont{J.}~\bibnamefont{Zheng}},
  \bibinfo{author}{\bibfnamefont{S.}~\bibnamefont{Bender}},
  \bibinfo{author}{\bibfnamefont{J.}~\bibnamefont{Armaitis}},
  \bibinfo{author}{\bibfnamefont{R.~E.} \bibnamefont{Troncoso}},
  \bibnamefont{and} \bibinfo{author}{\bibfnamefont{R.~A.} \bibnamefont{Duine}},
  \bibinfo{journal}{Phys. Rev. B} \textbf{\bibinfo{volume}{96}},
  \bibinfo{pages}{174422} (\bibinfo{year}{2017}),
  \urlprefix\url{https://link.aps.org/doi/10.1103/PhysRevB.96.174422}.

\bibitem[{\citenamefont{Brataas et~al.}(2000)\citenamefont{Brataas, Nazarov,
  and Bauer}}]{PhysRevLett.84.2481}
\bibinfo{author}{\bibfnamefont{A.}~\bibnamefont{Brataas}},
  \bibinfo{author}{\bibfnamefont{Y.~V.} \bibnamefont{Nazarov}},
  \bibnamefont{and} \bibinfo{author}{\bibfnamefont{G.~E.~W.}
  \bibnamefont{Bauer}}, \bibinfo{journal}{Phys. Rev. Lett.}
  \textbf{\bibinfo{volume}{84}}, \bibinfo{pages}{2481} (\bibinfo{year}{2000}),
  \urlprefix\url{https://link.aps.org/doi/10.1103/PhysRevLett.84.2481}.

\bibitem[{\citenamefont{Cherepanov et~al.}(1993)\citenamefont{Cherepanov,
  Kolokolov, and L'vov}}]{Cherepanov1993}
\bibinfo{author}{\bibfnamefont{V.}~\bibnamefont{Cherepanov}},
  \bibinfo{author}{\bibfnamefont{I.}~\bibnamefont{Kolokolov}},
  \bibnamefont{and} \bibinfo{author}{\bibfnamefont{V.}~\bibnamefont{L'vov}},
  \bibinfo{journal}{Phys. Rep.} \textbf{\bibinfo{volume}{229}},
  \bibinfo{pages}{81} (\bibinfo{year}{1993}),
  \urlprefix\url{https://www.sciencedirect.com/science/article/pii/037015739390107O}.

\bibitem[{\citenamefont{Hellman et~al.}(2017)\citenamefont{Hellman, Hoffmann,
  Tserkovnyak, Beach, Fullerton, Leighton, MacDonald, Ralph, Arena, D\"urr
  et~al.}}]{RevModPhys.89.025006}
\bibinfo{author}{\bibfnamefont{F.}~\bibnamefont{Hellman}},
  \bibinfo{author}{\bibfnamefont{A.}~\bibnamefont{Hoffmann}},
  \bibinfo{author}{\bibfnamefont{Y.}~\bibnamefont{Tserkovnyak}},
  \bibinfo{author}{\bibfnamefont{G.~S.~D.} \bibnamefont{Beach}},
  \bibinfo{author}{\bibfnamefont{E.~E.} \bibnamefont{Fullerton}},
  \bibinfo{author}{\bibfnamefont{C.}~\bibnamefont{Leighton}},
  \bibinfo{author}{\bibfnamefont{A.~H.} \bibnamefont{MacDonald}},
  \bibinfo{author}{\bibfnamefont{D.~C.} \bibnamefont{Ralph}},
  \bibinfo{author}{\bibfnamefont{D.~A.} \bibnamefont{Arena}},
  \bibinfo{author}{\bibfnamefont{H.~A.} \bibnamefont{D\"urr}},
  \bibnamefont{et~al.}, \bibinfo{journal}{Rev. Mod. Phys.}
  \textbf{\bibinfo{volume}{89}}, \bibinfo{pages}{025006}
  (\bibinfo{year}{2017}),
  \urlprefix\url{https://link.aps.org/doi/10.1103/RevModPhys.89.025006}.

\bibitem[{\citenamefont{Sadovnikov et~al.}(2018)\citenamefont{Sadovnikov,
  Grachev, Sheshukova, Sharaevskii, Serdobintsev, Mitin, and
  Nikitov}}]{PhysRevLett.120.257203}
\bibinfo{author}{\bibfnamefont{A.~V.} \bibnamefont{Sadovnikov}},
  \bibinfo{author}{\bibfnamefont{A.~A.} \bibnamefont{Grachev}},
  \bibinfo{author}{\bibfnamefont{S.~E.} \bibnamefont{Sheshukova}},
  \bibinfo{author}{\bibfnamefont{Y.~P.} \bibnamefont{Sharaevskii}},
  \bibinfo{author}{\bibfnamefont{A.~A.} \bibnamefont{Serdobintsev}},
  \bibinfo{author}{\bibfnamefont{D.~M.} \bibnamefont{Mitin}}, \bibnamefont{and}
  \bibinfo{author}{\bibfnamefont{S.~A.} \bibnamefont{Nikitov}},
  \bibinfo{journal}{Phys. Rev. Lett.} \textbf{\bibinfo{volume}{120}},
  \bibinfo{pages}{257203} (\bibinfo{year}{2018}),
  \urlprefix\url{https://link.aps.org/doi/10.1103/PhysRevLett.120.257203}.

\bibitem[{\citenamefont{Kim}(2012)}]{KIM2012217}
\bibinfo{author}{\bibfnamefont{J.-V.} \bibnamefont{Kim}}
  (\bibinfo{publisher}{Academic Press}, \bibinfo{year}{2012}),
  vol.~\bibinfo{volume}{63} of \emph{\bibinfo{series}{Solid State Physics}},
  pp. \bibinfo{pages}{217 -- 294},
  \urlprefix\url{http://www.sciencedirect.com/science/article/pii/B9780123970282000047}.

\bibitem[{\citenamefont{Jungwirth et~al.}(2018)\citenamefont{Jungwirth, Sinova,
  Manchon, Marti, Wunderlich, and Felser}}]{10754/627305}
\bibinfo{author}{\bibfnamefont{T.}~\bibnamefont{Jungwirth}},
  \bibinfo{author}{\bibfnamefont{J.}~\bibnamefont{Sinova}},
  \bibinfo{author}{\bibfnamefont{A.}~\bibnamefont{Manchon}},
  \bibinfo{author}{\bibfnamefont{X.}~\bibnamefont{Marti}},
  \bibinfo{author}{\bibfnamefont{J.}~\bibnamefont{Wunderlich}},
  \bibnamefont{and} \bibinfo{author}{\bibfnamefont{C.}~\bibnamefont{Felser}},
  \emph{\bibinfo{title}{The multiple directions of antiferromagnetic
  spintronics}} (\bibinfo{year}{2018}),
  \urlprefix\url{http://hdl.handle.net/10754/627305}.

\bibitem[{\citenamefont{Foerster and Maci{\`{a}}}(2019)}]{Foerster_2019}
\bibinfo{author}{\bibfnamefont{M.}~\bibnamefont{Foerster}} \bibnamefont{and}
  \bibinfo{author}{\bibfnamefont{F.}~\bibnamefont{Maci{\`{a}}}},
  \bibinfo{journal}{Journal of Physics: Condensed Matter}
  \textbf{\bibinfo{volume}{31}}, \bibinfo{pages}{190301}
  (\bibinfo{year}{2019}),
  \urlprefix\url{https://doi.org/10.1088%2F1361-648x%2Fab067c}.

\bibitem[{\citenamefont{Losby and Freeman}(2016)}]{losby2016spin}
\bibinfo{author}{\bibfnamefont{J.~E.} \bibnamefont{Losby}} \bibnamefont{and}
  \bibinfo{author}{\bibfnamefont{M.~R.} \bibnamefont{Freeman}},
  \emph{\bibinfo{title}{Spin mechanics}} (\bibinfo{year}{2016}),
  \eprint{1601.00674}.

\end{thebibliography}

\end{document}